\documentclass[a4paper,11pt]{article}
\pdfoutput=1 

\usepackage{jheppub} 

\usepackage[T1]{fontenc} 

\usepackage{latexsym,bm,amsmath,amssymb}
\usepackage{physics}
\usepackage{color}
\usepackage{tikz}
\usepackage{bbm}

\title{\boldmath Spherically-symmetric geometries in a matter reference frame as quantum gravity condensate microstates }

\author[a]{Daniele Oriti,}
\author[a]{Yi-Li Wang}

\affiliation[a]{Arnold Sommerfeld Center for Theoretical Physics, Ludwig-Maximilians-Universit\"at M\"unchen\\Theresienstra\ss e 37, 80333, M\"unchen, Germany, EU}
\emailAdd{daniele.oriti@physik.lmu.de}
\emailAdd{wang.yili@physik.uni-muenchen.de}

\abstract{Candidate microstates of a spherically symmetric geometry are constructed in the group field theory formalism for quantum gravity, for models including both quantum geometric and scalar matter degrees of freedom. The latter are used as a material reference frame to define the spacetime localization of the various elements of quantum geometry. By computing quantum geometric observables, we then match the candidate quantum microstates with a spherically symmetric classical geometry, written in a suitable matter reference frame.}

\begin{document} 
\maketitle
\flushbottom

\section{Introduction}
Spherically symmetric geometries, and Schwarzschild black holes in particular, play a special role in classical and quantum gravity. At the classical level, they are an important simplified model for large massive objects and for the result of gravitational collapse. They are crucial in quantum gravity because, first of all, any theory of quantum gravity is called to provide a microscopic derivation of the thermodynamical properties of black holes \cite{bekenstein,bekenstein2,hawking,hawking2,laws}, which become apparent already at a semi-classical level. Second, quantum gravity theories are expected to resolve the curvature singularity at the center of black holes and, more generally, to account for the Planckian regime reached in the last phase of black hole evaporation \cite{hawking}. These are important goals, but spherically symmetric geometries have also a more modest, yet crucial role, in quantum gravity. They represent the simplest yet highly non-trivial kind of spacetime geometries, after the homogeneous geometries used as cosmological backgrounds, and as such they are the first target (again, besides cosmological backgrounds) for the reconstruction of spacetime from fundamental quantum gravity degrees of freedom. In several quantum gravity formalisms, in fact, the fundamental structures are not directly spatiotemporal or geometric, and spacetime has to emerge from suitable collective dynamics \cite{condense,luca}. A well-defined and solid reconstruction procedure of spherically symmetric geometries is also a necessary ingredient, of course, for tackling the mentioned issues of quantum black hole physics from a fundamental quantum gravity perspective. \\

\noindent {\bf Reconstructing geometry in quantum gravity: a relational strategy} \\
In this contribution, we focus on this reconstruction task, in the context of a specific quantum gravity formalism in which space, time and geometry as we know them are emergent notions.  The broader issue is about how a continuum (and semiclassical) spacetime emerges from quantum gravity formalisms, together with the usual physical description in terms of quantum fields. This includes the emergence of curved spacetimes, thus gravity, but also flat spacetime without gravity (itself a highly excited state, from the non-perturbative quantum gravity perspective. This issue is common to basically all quantum gravity approaches; for some examples and discussions of results, physical and conceptual aspects, see for example \cite{Seiberg:2006wf,Padmanabhan:2014jta,Oriti:2013jga,Oriti:2018dsg}.
Specifically, we ask which effective continuum and classical geometry is compatible with a set of quantum gravity microstates defining a possible quantum microstructure of spherically symmetric spaces. The correspondence is necessarily based on matching the (average) value of specific geometric observables. Since any meaningful observable in classical and continuum gravity should be diffeomorphism invariant (at least under the assumption that general relativity (GR) is a good theoretical description), a first task is to define the needed geometric observables in a diffeomorphism invariant language. This is also necessary because all the structures that support diffeomorphism transformations in GR, i.e. the differentiable manifold and the continuum fields defined on it, and which the diffeomorphism invariance itself suggest to be redundant unphysical structures, may simply not be present in the fundamental theory. \\

A convenient, well-known strategy to define diffeomorphism invariant observables in classical and quantum gravity is in terms of relations between dynamical entities, each separately covariant under diffeos \cite{partial1,partial2}. Physically, this usually corresponds to singling out a subset of the dynamical degrees of freedom and use them to define a (physical, as opposed to a coordinate) reference frame, in terms of which one then define the localization `in time and space' and the evolution of the remaining dynamical degrees of freedom of the theory. A common choice for such physical reference frame is suitable matter fields, with the simplest option being represented by scalar fields, one for each spacetime direction. Since a physical reference frame would remain a dynamical and quantum system, it would in general define space and time localization (in the same way a coordinate frame does) only approximately \cite{Hoehn:2019fsy,luca}, i.e. in the approximation in which it behaves like an ideal set of rods and clock (e.g. having negligible energy-momentum tensor and thus backreaction on the geometry). \\

A first main challenge we will tackle in the following is to define quantum gravity microstates that could represent candidates for the microstructure of spherically symmetric geometries and include as well the dynamical degrees of freedom of a suitable physical reference frame. It is in such physical reference frame that an effective continuum geometry can be reconstructed, on the basis of a few relevant geometric observables. This will represent an improvement over the quantum microstates constructed in \cite{bhlett,bhc} for the same purpose, but use additional parameters of the wavefunctions instead of dynamical degrees of freedom to identify localization and support the interpretation in terms of spherical symmetry. We will also show that this extension of the quantum microstates maintains the main properties (notably, holographic character and entropy area law) of the previous construction.\\

The second main challenge will be to identify in which precise reference frame the reconstructed continuum geometry has been written, that is, which continuum fields can be put in correspondence with the reference frame data that are present in the fundamental quantum microstates. Ideally, this information would be encoded in (or at least strongly constrained by) the (quantum) dynamical equations satisfied by our quantum gravity microstates, and the task would be then to check the compatibility between the quantum microstates themselves and a desired effective continuum geometry, seen as a test of physical viability for the same quantum microstates. However, no such information is possessed, in our case, since our construction will be purely kinematical. That is, we do not know if the quantum microstates we construct solve the fundamental quantum dynamics of a GFT model for 4d quantum gravity coupled to specific matter degrees of freedom (which would then be used to define the physical reference frame). This is the main limitation of our results. The other main limitation is that we have control only over a limited set of quantum geometric observables, and this impacts the extent to which we can check the compatibility of our quantum microstates with their interpretation as spherically symmetric quantum geometries.\\

Our procedure still illustrates the key challenges of any reconstruction of an effective continuum geometry from ` pre-geometric'  quantum gravity microstates, and thus its significance goes beyond the specific formal context we work with.\\


\noindent {\bf The GFT formalism for quantum gravity and spherically symmetric states}\\
The quantum gravity formalism we work with is Group Field Theory (GFT) \cite{gft1,gft2,gft3,gft4,gft5,gft6}. It can be understood from different perspectives. GFTs are a field-theoretic and combinatorial generalization of random matrix models for 2d quantum gravity to higher dimensions, in the sense that their perturbative interaction processes (Feynman diagrams) are dual to d-dimensional cellular complexes. From this point of view they belong, together with the purely combinatorial random tensor models \cite{Gurau:2011xp,Rivasseau:2016wvy}, to the general class of models labeled Tensorial Group Field Theories (TGFT), characterized by combinatorially non-local interactions that are responsible for the topology of the Feynman diagrams. The control over the topology associated to quantum microstates and amplitudes, provided by the tools of TGFTs in general, will be crucial for our construction, since they are used to define the needed sum over spherically symmetric lattices associated to the relevant quantum microstates. \\

Beside their combinatorial aspects, GFTs are proper field theories, specifically based on quantum fields on group manifolds (or their corresponding Lie algebras/representation spaces), and thus endowed with rich algebraic data. These algebraic data provide a quantum geometric interpretation to the combinatorial structures of GFT models, that is the quanta of the GFT fields, the quantum microstates and observables, and the GFT Feynman amplitudes. At this level, GFT models connect with canonical loop quantum gravity \cite{lqgintro1,lqgintro2,lqgintro3,lqgintro4,lqgintro5,lqgintro6}, with which they share the basic features of quantum microstates (with a basis given by spin networks), and of which GFT can be seen as a second quantized reformulation (for GFT models based on the $SU(2)$ group manifold) \cite{2nd}, and with spin foam models \cite{sf} and lattice gravity path integrals, whose amplitudes coincide with the GFT Feynman amplitudes (for given models) \cite{gft1,gft2,gft3,gft4,gft5,gft6}. The quantum geometric data are also crucial for our construction because they guide our interpretation of the states, first of all, but also because they allow for the definition of a quantum many-body Fock space of states, which we work with at the quantum level.\\

We introduce here the basic ingredients of the GFT formalism, limiting ourselves to the ones we use in the following, and to quantum geometric models based on $SU(2)$ data. For more extensive introductions, we refer to the literature \cite{gft1,gft2,gft3,gft4,gft5,gft6}.\\

A GFT is a quantum field theory of fields on group manifolds, whose quanta are interpreted, for quantum geometric models, as discrete (usually simplicial) building blocks of quantum space (more generally, codimension-1 quantum geometries). For 4d quantum gravity, GFT quanta are pictured as $3$-simplices (tetrahedra) dual to $4$-valent spin network (SNW) vertices. Specific quantum microstates can be then associated to simplicial complexes or extended 4-valent graphs by gluing together such quanta, with the gluing effected by (maximal) entanglement of the individual quanta (with respect to the quantum degrees of freedom associated to faces in the two simplices corresponding to the different GFT quanta). \\

The dynamical degrees of freedom corresponding to scalar fields $\phi_i$ can be added, in a way that matches how continuum scalar fields are discretized on the simplicial complexes dual to the GFT states and processes; scalar field values in fact end up being associated to SNW vertices.  In this article, the GFT quantum dynamics and thus the processes involving such simplicial objects, endowed with both quantum geometric and scalar field degrees of freedom, will not be studied. Therefore, the dependence of the GFT field on such data and their coupling will only be constrained by kinematical and geometrical considerations. Moreover, the matter degrees of freedom will be analysed only as providing a physical reference frame for the definition of relational, diffeomorphism invariant geometric observables, rather than for their physical consequences. Localizing a point in four-dimensional spacetime needs four scalar fields, one as a clock while three as rods, so we will work with GFT states for quantum geometry coupled to four scalar fields. Since we are interested in quantum microstates that correspond to macroscopic spherically symmetric geometries, we take the scalar fields to define spherical frames, with $\phi_0$ binge the clock and $\phi_1$ being the `radial' rod, with $\phi_2$ and $\phi_3$ determining points on a 2-sphere. Then the complex GFT field is a map $\varphi: SU(2)^4 \times \mathbb{R}^4 \to \mathbb{C}$. 
Letting $\phi =(\phi_0,\phi_1,\phi_2,\phi_3)$ be the scalar field values and $g=(g_1,g_2,g_3,g_4)$ the group elements encoding discretized connection data in each 3-simplex, the GFT field corresponds to creation/annihilation operators $\hat{\varphi}^{\dagger}(g,\phi)$ and $\hat{\varphi}(g,\phi)$ satisfying 
\begin{eqnarray}
&&[\hat{\varphi}(g,\phi),\hat{\varphi}^{\dagger}(g',\phi')]\nonumber\\
&=&\delta^4(\phi-\phi')\int_{SU(2)}d\gamma\prod_{i=1}^4\delta(g_i'\gamma g_i^{-1})
\equiv\Delta_R(g,g')\delta^4(\phi-\phi), \label{commutator}
\end{eqnarray}
where $\delta(g)$ is the delta function over $SU(2)$, and commutators between creation or annihilation operators vanish. 
These operators, like the GFT field, are invariant under the diagonal action of $SU(2)$ on all the group arguments:
\begin{equation}
\hat{\varphi}(g k,\phi)=\hat{\varphi}(g,\phi),
\end{equation}
where $k\in SU(2)$; this is the origin of the diagonal integration on the r.h.s of \eqref{commutator}. \\

The creation/annihilation operators act on a Fock space, creating/destroying individual tetrahedra (equivalently, spin network vertices), with the Fock vacuum $\ket{0}$ representing a state of no building block. A generic n-body Fock state is thus built from field operators
\begin{equation}
\ket{\psi}=\int dg_i \psi(g_1,...,g_n)\prod_{i=1}^n\hat{\varphi}^{\dagger}(g_i)\ket{0},
\end{equation} 
and among such n-body states one can identify those that an be associated to connected graph structures, or equivalently 3d simplicial complexes) looking at the entanglement properties of the wavefunction $\psi(g_i)$ \cite{2nd,Colafranceschi:2020ern,Colafranceschi:2021acz,Chirco:2021chk}. Such n-body states coincide with the quantum microstates of spatial geometry in loop quantum gravity (up to embedding information, absent in GFT, and now organized in a different Hilbert space, the GFT Fock space), so that GFT can be understood as a second quantization of the latter. The quantum dynamics of GFT states, in the standard QFT language, is encoded in operator equations 
\begin{equation}
\frac{\delta S[\hat{\varphi},\hat{\varphi}^{\dagger}]}{\delta \bar{\varphi}(g)}\ket{\psi}=0,
\end{equation}
where $S[\hat{\varphi},\hat{\varphi}^{\dagger}]$ is an action functional. It can also be expressed in terms of Schwinger-Dyson equations for N-point functions, extracted from a path integral formulation.\\ 


The article is organised as follows:
Section $2$ introduces the GFT quantum microstates of a spherically symmetric continuum quantum geometry in a material reference frame, followed by the calculation of the expectation value of relational geometric observables and the investigation on some interesting properties of these states, in section $3$. In section $4$ we identify which material reference systems would produce the same relational observables in a continuum and classical spherically symmetric geometry; by matching the results with those obtained in the quantum gravity context, we find the restrictions on the GFT states (notably, the single vertex wavefunction) that ensure compatibility of the two sets of results, and thus the viability of the spacetime interpretation of the same quantum microstates. Section $5$ presents a summary and outlook.


\section{Spherically symmetric GFT condensate states}
\begin{figure}[tbp]
	\centering 
	\includegraphics[width=.45\textwidth]{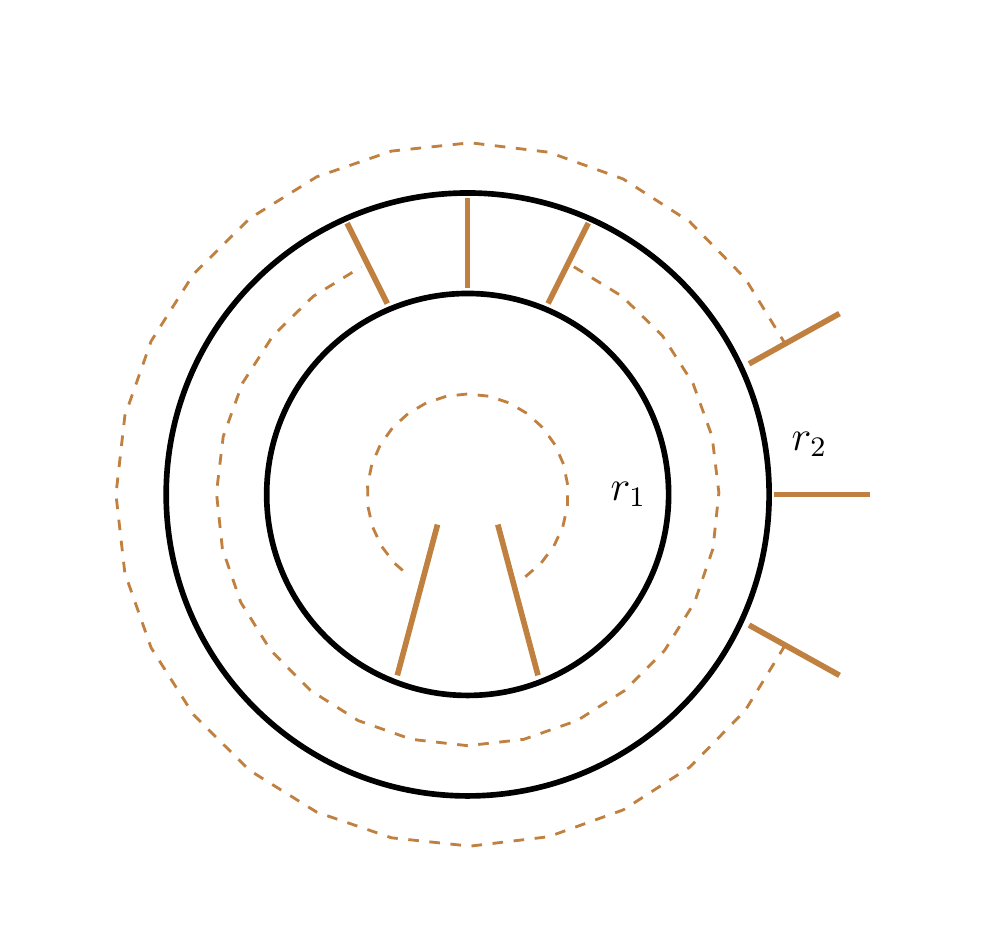}
	\caption{\label{shell} A spherically symmetric spatial geometry is obtained through a foliation into homogeneous shells. The space foliation is achieved through gluing the outer boundary on shell $r_1$ and inner boundary on shell $r_2$. }
\end{figure}

We now detail the construction of GFT quantum microstates for spherically symmetric continuum geometries, extending the construction in \cite{condensate,bhlett,bhc} by the inclusion of scalar field degrees of freedom, first, and then by the use of special single-vertex wavefunctions peaked on given values of such scalar field variables, defining a relational frame. These quantum geometries can be seen as corresponding special kinds of GFT condensates. 
Some of the elements in the construction have been previously applied for extracting an effective cosmological dynamics \cite{con1,condense,effective}, for GFT states associated to continuum homogeneous spatial geometries. 
The idea is to consider spherically symmetric 3-spaces as foliated into thin concentric shells \cite{bhc,bhlett}, as illustrated by figure \ref{shell}. Each shell is endowed with a homogeneous quantum geometry, and labelled by a `radial' parameter $r$. The outer boundary of $r_1$-shell is then glued with the inner boundary of $r_2$-shell to form the full spatial foliation. \\

A key aspect of the construction is that the constructed quantum microstates incorporate a sum over graph/complexes (with the appropriate topology), and thus do not depend on any single lattice nor, a priori, on any truncation to finite number of quantum geometric degrees of freedom, since the number of vertices in the graphs summed over can be extended to arbitrary values; this is what justifies the interpretation in terms of continuum spatial geometries.
Of course, the assignment of quantum geometric data should then be shown to be compatible with the topological structure, for the intended interpretation to be viable.\\

Thus the first task is to define quantum microstates of each shell, which can be consistently understood as endowed with a continuum and semi-classical homogeneous geometry \cite{condense}.
The details of the construction is given below. The main features are
\begin{itemize}
	\item [i.] Spherical topology: the full shell states are defined by superpositions of quantum microstates associated to graphs dual to spherically symmetric simplicial complexes with shell topology, and increasing number of 3-simplices;
	\item [ii.] Homogeneity: the quantum geometrical information associated to every GFT quanta forming these shell states is the same; in other words, each GFT quantum is assigned the same (condensate) wavefunction; this is a microscopic (as opposed to coarse-grained) counterpart of spatial homogeneity in the sense that such assignment implies the reconstruction of the same geometric information at the `locus' of the dual simplex, from the quantum microstate, upon embedding of the corresponding simplicial complex into some auxiliary 3-manifold \cite{condense};
	\item [iii.] Reference frame peakedness: the condensate wavefunction assigned to each and every vertex/simplex is sharply peaked around appropriate values of the (matter) degrees of freedom chosen to define a physical reference frame; that is, these internal degrees of freedom provide a good set (i.e. up to small quantum fluctuations) of clock and rods (they can be seen also as defining a local embedding into $\mathbb{R}^4$);
	\item [iv.] Near-flatness: the geometry inside the simplices corresponding to GFT quanta is nearly flat, to ensure a better geometric interpretation in a continuum embedding (this condition, in fact, although assumed in the original construction, plays little role in what has been done so far with these quantum microstates);
	\item [v.] Continuum approximation: the linear superposition defining the quantum microstates is extended to graphs/simplicial complexes with a very large number of vertices/simplices (potentially, up to infinity); this takes care of the combinatorial aspect of the continuum approximation; we then need to take care of the matching of geometric information;
	\item [vi.] Semi-classicality: the (relative) quantum fluctuations of geometric observables, around the expectation values that we will determine and that should match the ones of classical geometry, should be small; for the observables we consider here, which are additive in the GFT number of quanta, this condition is ensured already by the use of largely populated quantum microstates.
\end{itemize}

As anticipated above, for each spherical shell, some simple geometric operators will be defined and their expectation value computed; this can be done for each shell state. Then, the gluing of shells needs to be performed to obtain the final quantum microstates for spherically symmetric quantum geometries. Before proceeding, in the next section, to the matching with classical continuum geometries in terms of the chosen observables, and the task of identifying the appropriate material reference frame in which the matching can be obtained, we will summarize some more features of the quantum microstates, following the steps in \cite{condense}.\\

\subsection{The seed state and refinement}
This subsection outlines the combinatorial aspects of the construction; techniques from the random tensor models literature will be crucial \cite{Gurau:2011xp,Rivasseau:2016wvy}. We will define linear combinations of states for spherically symmetric shells, each corresponding to the triangulation of a spherical shell with two boundaries, thus the topology $S^2\times [0,1]$. These triangulations are constructed by acting on an initial seed state 
 $\ket{\tau}$ with appropriate refinement operations, which increase the number of simplices in the triangulation without changing the overall topology. The simplicial topology can be encoded in a bipartite colored 4-graph, using the standard tools of TGFTs (and random tensor models \cite{Gurau:2011xp,Rivasseau:2016wvy}).\\

As a simple triangulation of a spherical shell, the seed state can be represented graphically as
\begin{equation}\label{seed0}
\begin{tikzpicture}
\draw [dashed] (-1.5,0)--(0,0);
\draw (0,0)--(5,0);
\draw [dashed] (5,0)--(6.5,0);
\draw (0,0)--(0,-1);
\draw (0,-1)--(5,-1);
\draw (5,-1)--(5,0);
\draw (0,0)--(0,2.5);
\draw (0,2.5)--(5,2.5);
\draw (5,2.5)--(5,0);
\draw (1,2.5)--(1,3);
\draw (1,3)--(4,3);
\draw (4,3)--(4,2.5);
\draw (1,2.5)--(1,1);
\draw (1,1)--(4,1);
\draw (4,1)--(4,2.5);
\draw (1.5,1)--(1.5,0.5);
\draw (1.5,0.5)--(3.5,0.5);
\draw (3.5,0.5)--(3.5,1);
\draw [dashed] (1.5,1)--(1.5,1.5);
\draw [dashed] (3.5,1)--(3.5,1.5);
\draw [fill] (0,0) circle [radius=0.1];
\draw [fill=white] (1,2.5) circle [radius=0.1];
\draw [fill] (1.5,1) circle [radius=0.1];
\draw [fill=white] (5,0) circle [radius=0.1];
\draw [fill] (4,2.5) circle [radius=0.1];
\draw [fill=white] (3.5,1) circle [radius=0.1];
\node [below] at (-1,-0.2) {1};
\node [left] at (0,1.5) {2};
\node [below] at (2.5,0) {3};
\node [below] at (2.5,-1) {4};
\node [below] at (6,-0.2) {1};
\node [right] at (5,1.5) {2'};
\node [above] at (2.5,3) {1''};
\node [below] at (2.5,2.5) {4''};
\node [left] at (1,1.5) {3''};
\node [right] at (4,1.5) {3'''};
\node [above] at (1.5,1.5) {4};
\node [above] at (3.5,1.5) {4};
\node [above] at (2.5,1) {1'''};
\node [above] at (2.5,0.4) {2'''};
\end{tikzpicture}
\end{equation}
Each four-valent vertex corresponds to a simplicial building block, and it is associated with a colour $t=\{B(lack),W(hite)\}$. These new labels amount to working with multiple \lq atomic species\rq, each corresponding to a different set of field operators. Every link is labeled by a colour from $1$ to $4$ (these were already assumed in the definition of the field operators), and only vertices of different colours can be connected. This state contains three parts: the bulk, the outer boundary and the inner boundary, corresponding to vertices without open links, with one open link of colour $1$ and one open link of colour $4$, respectively. To connect shells $r_1$ and $r_2$, in the next step, as in figure \ref{shell}, the open links on the outer boundary of $r_1$ must share the same colour of those on the inner boundary of $r_2$. Moreover, to distinguish these three parts of each shell, a new label $s$ is used, representing outer boundary, bulk and inner boundary: $s=\{+,0,-\}$. These additional labels, contrary to the previous ones, are associated to the condensate wavefunctions that will correspond to different created/annihilated vertices, thus do not correspond to different fields. \\

The refinement operator $\hat{\mathcal{M}}_{t,s}$, to be defined functionally in the next subsection, eliminates one vertex in the graph and adds three new ones, connected to form a simple melonic graph; in other words, it corresponds to a dipole insertion, in the tensor model language. This move preserves the topology. \\

Each layer of a shell has its own associated refinement operators, which of course have to act in parallel, in order to preserve the connectivity. 
In a graphical representation, one has, for instance, a refinement on outer boundary
\begin{equation}
\begin{tikzpicture}
\node at (-2,0) {$\hat{\mathcal{M}}_{W+}:$};
\draw (-1,0)--(1,0);
\draw (0,-1)--(0,1);
\draw [fill=white] (0,0) circle [radius=0.1];
\node [right] at (0,0.5) {1};
\node [above] at (-0.5,0) {2};
\node [left] at (0,-0.5) {3};
\node [below] at (0.5,0) {4};
\draw [->] (1.5,0)--(2,0);
\draw (3,0)--(7,0);
\draw (4,-1)--(4,1);
\draw (5,-1)--(5,1);
\draw (6,-1)--(6,1);
\draw (4,-1)--(5,-1);
\draw [fill=white] (4,0) circle [radius=0.1];
\draw [fill] (5,0) circle [radius=0.1];
\draw [fill=white] (6,0) circle [radius=0.1];
\node [left] at (4,0.8) {1''};
\node [left] at (5,0.8) {1'};
\node [left] at (6,0.8) {1};
\node [below] at (3,0) {2};
\node [above] at (4.5,-1) {3'};
\node [above] at (4.5,0) {4'};
\node [above] at (5.5,0) {2'};
\node [above] at (7,0) {4};
\node [above] at (6.2,-1) {3};
\node at (7.2,-1) {,};
\end{tikzpicture}
\end{equation}
and
\begin{equation}
\begin{tikzpicture}
\node at (-2,0) {$\hat{\mathcal{M}}_{B+}:$};
\draw (-1,0)--(1,0);
\draw (0,-1)--(0,1);
\draw [fill] (0,0) circle [radius=0.1];
\node [right] at (0,0.5) {1};
\node [above] at (-0.5,0) {4};
\node [left] at (0,-0.5) {3};
\node [below] at (0.5,0) {2};
\draw [->] (1.5,0)--(2,0);
\draw (3,0)--(7,0);
\draw (4,-1)--(4,1);
\draw (5,-1)--(5,1);
\draw (6,-1)--(6,1);
\draw (4,-1)--(5,-1);
\draw [fill] (4,0) circle [radius=0.1];
\draw [fill=white] (5,0) circle [radius=0.1];
\draw [fill] (6,0) circle [radius=0.1];
\node [left] at (4,0.8) {1''};
\node [left] at (5,0.8) {1'};
\node [left] at (6,0.8) {1};
\node [below] at (3,0) {4};
\node [above] at (4.5,-1) {3'};
\node [above] at (4.5,0) {2'};
\node [above] at (5.5,0) {4'};
\node [above] at (7,0) {2};
\node [above] at (6.2,-1) {3};
\node at (7.2,-1) {.};
\end{tikzpicture}
\end{equation}
The operations are similar when acting on other layers, so we do not illustrate the others here. Thus, each layer would consist of $(2n+2)$ vertices after $n$ refinement moves.\\

Now we move to the GFT ladder operators which define the relevant condensate states and the refinement operators acting on them.

\subsection{Ladder operators and coherent peaked states}


To ensure that the homogeneity encoded in the assignment of equal wavefunctions to every vertex (i.e. the condition of (microscopic) homogeneity of the quantum geometry) is preserved under refinement, we need this to be assigned to any newly created vertex, and removed by the annihilation of the existing one. This is accomplished by the definition of new creation/annihilation operators `dressed' by the condensate wavefunction itself, and function of appropriate `gluing variables':
\begin{eqnarray}\label{condensate}
\hat{\sigma}_{t}(h_I)=\int dg d\phi \sigma(h_I g_I,\phi)\hat{\varphi}_{t}(g_I,\phi),\\
\hat{\sigma}^{\dagger}_{t}(h_I)=\int dg_I d\phi_v \overline{\sigma(h_Ig_I,\phi)}\hat{\varphi}^{\dagger}_{t}(g_I,\phi).
\end{eqnarray}
The group elements $h=(h_1,h_2,h_3,h_4)$ are auxiliary variables at the endpoints of links, ensuring that the connectivity is realised through convolution (thus entanglement of the vertex states). 
\\

To use these condensate field operators to create a shell, as anticipated and illustrated in fig. \ref{shell}, one label shells according to their `locations'. 
Following recent literature on the cosmological dynamics of GFT condensates, we introduce these labels by using peaked condensate wavefunctions \cite{luca,luca2}
\begin{equation}\label{full}
\sigma_{\tau,u,\theta,\chi}(g;\phi)\equiv\eta_{\epsilon_0}(\phi_0-\phi_{\tau},\pi_{\tau})\eta_{\epsilon_1}(\phi_1-\phi_{u},\pi_u)\eta_{\epsilon_2}(\phi_2-\phi_{\theta},\pi_{\theta})\eta_{\epsilon_3}(\phi_3-\phi_{\chi},\pi_{\chi})\varsigma(g;\phi),
\end{equation}
where
\begin{equation}
\eta_{\epsilon_x}\equiv\mathcal{N}_{\epsilon_x}\exp(-\frac{(\phi_x-x)^2}{2\epsilon_x})\exp(i\pi_x(\phi_x-x))
\end{equation}
is a peaking function, here chosen to be a gaussian, and $\pi_x$ is the conjugate momentum with respect to $\phi_x$. The constant $\mathcal{N}_{\epsilon_x}^2=(\pi\epsilon_x)^{-1/2}$ is a normalisation factor. The function $\varsigma$ is polynomial in the scalar field dependence, so to not spoil peakedness properties, but otherwise arbitrary, and should be then determined by the dynamics or by further geometric restrictions. A small deviation from peak values is ensured by the requirement $\epsilon_x\ll 1$. The interpretation is that we have a condensate wavefunction approximately \lq at the location corresponding to the peak values of the scalar fields\rq, used as a material reference frame.
For instance, an operator 
\begin{equation}
\hat{\sigma}^{\dagger}_t(h_v;\{\phi_{\tau},\phi_{u},\phi_{\theta},\phi_{\chi}\})=\int dg_v d\phi_v\overline{\sigma_{\tau,u,\theta,\chi}(h_vg_v,\phi_v)}\hat{\varphi}^{\dagger}(g_v,\phi_v)
\end{equation}
produces, with small fluctuations, a tetrahedron at $(\phi_0=\phi_{\tau},\phi_1=\phi_u,\phi_2=\phi_{\theta},\phi_3=\phi_{\chi})$. Because we are dealing with a spherically symmetric static object, the wavefunction should depends only on the `radial' $\phi_1$. That is, all vertices in a given shell look the same, but vertices on different shells do not, in terms of their quantum geometric data. Denoting $\phi_1$ by $\phi$, 
\begin{eqnarray}
&&\hat{\sigma}_{t}^{\dagger}(h;\phi_u)\equiv\int dg d\phi \overline{\sigma_u(hg,\phi)}\hat{\varphi}^{\dagger}_t(g,\phi),\\
&&\hat{\sigma}_{t}(h;\phi_u)\equiv\int dg d\phi \sigma_u(hg,\phi)\hat{\varphi}_t(g,\phi),
\end{eqnarray} 
where 
\begin{equation}\label{peak}
\sigma_u(g;\phi)=\eta_{\epsilon}(\phi-\phi_u,\pi_u)\varsigma(g;\phi).
\end{equation}
\\

Thanks to these coherent peaked states and the \lq localization information\rq, we have a more straightforward control over geometric information, and, we would argue, a more physically sensible one. In the original constructions \cite{bhc,condensate}, there are two labels $r$ and $s$ characterising the condensate wavefunction $\sigma_{r,s}$. Now these localization labels corresponds to average values of physical (matter) degrees of freedom, according to the relational strategy for the definition of spacetime localized observables in classical and quantum gravity \cite{Hoehn:2019fsy,luca}. 
\\

Since the states of different layers must be orthogonal, to avoid interference effects that would put into question a semiclassical reconstruction of continuum geometry, the ladder operators with different peaks should commute. Let us briefly analyze the conditions for peaks in order to generate distinguishable shells. According to the definition of peaking states, what contributes is the Gaussian function $\exp((\phi-\phi_u)^2/2\epsilon)$. As $\epsilon$ is small, one approximates the result as the one measured at $\phi=\phi_a$. Since $\epsilon$ is small but finite, the peaking function actually has a width of $6\sqrt{\epsilon}$, counting the contribution from $\phi=\phi_u-3\sqrt{\epsilon}$ to $\phi=\phi_u+3\sqrt{\epsilon}$. Then peaks should not overlap with each other, so they should be chosen to differ by an amount bounded as
\begin{equation}\label{constraint}
|\phi_{u_1}-\phi_{u_2}|\geq 6\sqrt{\epsilon},
\end{equation}
which then ensures
\begin{equation}\label{orthogonal}
[\hat{\sigma}(h;\phi_{u_1}),\hat{\sigma}^{\dagger}(h,\phi_{u_2})]\simeq 0.
\end{equation}

Assuming the above conditions, we study the commutator between $\hat{\sigma}_t(h_v;\phi_u)$ and $\hat{\sigma}^{\dagger}_t(h_w;\phi_{u'})$, and identify the conditions to ensure the homogeneity of the shell quantum geometry. In the following, we use a label $v$ to indicate variables (thus, conditions) that correspond to individual vertices.\\

First one performs a Peter-Weyl decomposition on the wavefunction \cite{gplqg}
\begin{equation}\label{pw}
\varsigma(g_v,\phi_v)=\sum_{{j},l_L,l_R}\iota^{j_1j_2j_3j_4l_L}_{m_1m_2m_3m_4}\iota^{j_1j_2j_3j_4l_R}_{n_1n_2n_3n_4}\varsigma^{j_1j_2j_3j_4l_Ll_R}(\phi_v)\prod_{i=1}^{4}D_{m_in_i}^{j_i}(g_{v,i}),
\end{equation}
where the intertwiner
\begin{equation}
\iota_{m_1m_2m_3m_4}^{j_1j_2j_3j_4}=\sum_{m,m'}C_{m_1m_2m}^{j_1j_2l}C_{m_3m_4m'}^{j_3j_4l'}C_{mm'0}^{ll'0},
\end{equation}
with $C^{jj_1j_2}_{mm_1m_2}$ the Clebsch-Gordon coefficient. With the convention
\begin{equation}
\int dg D_{mn}^{j}(g)\overline{D_{m'n'}^{j'}(g)}=\frac{1}{d_j}\delta_{mm'}\delta_{nn'}\delta_{jj'},
\end{equation}
where $d_j=2j+1$, and
\begin{equation}
\sum_{m}\iota_{m_1m_2m_3m_4}^{j_1j_2j_3j_4l}\iota_{m_1m_2m_3m_4}^{j_1j_2j_3j_4l'}=\delta^{l,l'}n(j_1,j_2,j_3,j_4,l),
\end{equation}
when one assumes
\begin{eqnarray}
&&\varsigma^{j_1j_2j_3j_4l_Ll_R}(\phi_u)=\delta_{l_L,l_R}\varsigma^{j_1j_2j_3j_4l_L}(\phi_u),
\end{eqnarray}
\begin{eqnarray}\label{assumption}
&&\varsigma^{j_1j_2j_3j_4l_R}(\phi_u)\overline{\varsigma^{j_1j_2j_3j_4l_R}(\phi_u)}=
\frac{1}{\tilde{\varsigma}(\phi_u)}\frac{d_{j_1}^2d_{j_2}^2d_{j_3}^2d_{j_4}^2}{n(j_1,j_2,j_3,j_4,l_R)},
\end{eqnarray}
the commutator reads
\begin{equation}\label{commutation}
[\hat{\sigma}_{t}(h_v;\phi_u),\hat{\sigma}_{t'}^{\dagger}(h_w;\phi_{u'})]
=\delta_{u,u'}\delta_{t,t'}\frac{1}{\tilde{\varsigma}(\phi_u)}\int_{SU(2)}d\gamma\prod_{i=1}^4\delta(\gamma h_{v_i}h_{w_i}^{-1})
\equiv\delta_{r,r'}\delta_{t,t'}\frac{1}{\tilde{\varsigma}(\phi_u)}\Delta_L(h_v,h_w).
\end{equation}
The assumption (\ref{assumption}) is rather strong, and it is equivalent to separating the variable $\phi$ in the original wavefunction. However, this assumption can be shown to be necessary to obtain the desired commutator (\ref{commutation}).\\ 

A proper geometric interpretation, and the subsequent matching of our quantum microstates with semiclassical continuum geometries, requires the calculation of quantum geometric observables. Such matching imposes, as we will see, non-trivial conditions on the matter reference frame we chosen to define relational geometric observables. At the quantum level, it also imposes conditions on the condensate wavefunction characterizing our quantum microstates. 
Indeed, by computing quantities such as areas and volume of a shell, we will see how these geometric observables depend on $\sigma$. Their expectation values will have to match the classical ones of a spherically symmetric geometry, e.g. Schwarzschild black hole, for our states to be good candidate microstates of the same geometry. We now turn to the computation of such observables. 
\\ 

\subsection{Shell states and extended $(1+1)$-body operator}
Assume that a shell consists of three types of tetrahedra governed by wavefunctions with slightly different peaks. Letting $6\sqrt{\epsilon}\leq\delta \phi_u$, one defines
\begin{eqnarray}
&&\hat{\sigma}_{u,t0}^{\dagger}(h)\equiv\hat{\sigma}_{t}^{\dagger}(h;\phi_u),\\
&&\hat{\sigma}^{\dagger}_{u,t+}(h)\equiv\hat{\sigma}_{t}^{\dagger}(h;\phi_u+\delta \phi_u),\\
&&\hat{\sigma}^{\dagger}_{u,t-}(h)\equiv\hat{\sigma}_{t}^{\dagger}(h;\phi_u-\delta \phi_u).
\end{eqnarray}
Additionally, $\tilde{\varsigma}(\phi_u)$ is replaced by $\tilde{\varsigma}_{us}$ for convenience, where
\begin{eqnarray}
&&\tilde{\varsigma}_{u0}=\tilde{\varsigma}(\phi_u),\\
&&\tilde{\varsigma}_{u+}=\tilde{\varsigma}(\phi_u+\delta \phi_u),\\
&&\tilde{\varsigma}_{u-}=\tilde{\varsigma}(\phi_u-\delta \phi_u).
\end{eqnarray}
A $u$-shell means a state whose bulk is built by $\hat{\sigma}^{\dagger}_t(h;\phi_u)$, and its seed state (\ref{seed0}) can be obtained by six condensate operators acting on the Fock vacuum:
\begin{eqnarray}\label{seed}
\ket{\tau_u}&\equiv&\int dg^{10}\hat{\sigma}_{u,B+}^{\dagger}(e,g_2,g_3,g_4)\hat{\sigma}_{u,W+}^{\dagger}(e,g_2',g_3,g_4)\hat{\sigma}_{u,B0}^{\dagger}(g_1'',g_2',g_3''',g_4'')\nonumber\\
&&\hat{\sigma}_{u,W0}^{\dagger}(g_1'',g_2,g_3'',g_4'')\hat{\sigma}_{u,B-}^{\dagger}(g_1'''',g_2'''',g_3'',e)\hat{\sigma}_{u,W-}^{\dagger}(g_1'''',g_2'''',g_3''',e)\ket{0}.
\end{eqnarray}
Similarly, the refinement on white vertices at an outer boundary $\phi_u$ is 
\begin{eqnarray}
\hat{\mathcal{M}}_{u,W+}&\equiv& \tilde{\varsigma}^{2}_{us}\int dk_2 dk_3 dk_4 dh_{4'}dh_{3'}dh_{2'}\hat{\sigma}^{\dagger}_{u,W+}(e,k_2,h_{3'},h_{4'})\hat{\sigma}^{\dagger}_{u,B+}(e,h_{2'},h_{3'},h_{4'})\nonumber\\
&&\hat{\sigma}^{\dagger}_{u,W+}(e,h_{2'},k_3,k_4)\hat{\sigma}_{u,W+}(e,k_2,k_3,k_4),
\end{eqnarray}
and the others can be expressed in the same manner. Thus a shell state is
\begin{equation}\label{single shell}
\ket{\Psi_u}=F(\hat{\mathcal{M}}_{u,Bs},\hat{\mathcal{M}}_{u,Ws})\ket{\tau_u},
\end{equation}
with $F$ a generic functional of refinement operators, whose polynomial order defines the truncation in the number of quantum geometric degrees of freedom. One can assume each layer to have the same number $n$ of the black and white tetrahedra. The quantum microstates defined in this way can be interpreted as an (approximately) continuum geometry when the number of tetrahedral are very large (up to $n\to\infty$).  \\ 

A few more details are needed, for the following computation. 
From now on, $\ket{\Gamma_n}$ represents a shell with $6n$ vertices as a general linear combination of $\ket{\Psi_u}$, constructed from the seed state through $(n-1)$ refinements on each layer:
\begin{eqnarray}
\ket{\Gamma_n(\phi_u)}&\equiv&\frac{1}{(n!)^3}\prod_{s=\{+,0,-\}}\tilde{\varsigma}_{us}^{n-1}
\int\prod_{i=1}^{6n} dh_i J_{\Gamma}(h_1,...,h_{6n})
\hat{\sigma}_{u,B+}^{\dagger}(h_1)...\hat{\sigma}_{u,W+}^{\dagger}(h_{n+1})...
\hat{\sigma}_{u,B0}^{\dagger}(h_{2n+1})\nonumber\\
&&...\hat{\sigma}_{u,W0}^{\dagger}(h_{3n+1})...
\hat{\sigma}_{u,B-}^{\dagger}(h_{4n+1})...\hat{\sigma}_{u,W-}^{\dagger}(h_{5n+1})...\hat{\sigma}_{u,W-}^{\dagger}(h_{6n})\ket{0},
\end{eqnarray}
where $J_{\Gamma}$ combines Dirac delta functions to ensure the links in the graph are appropriately connected. 
\\

The next step is to compute the expectation value of geometric observables. A generic $(1+1)$-body operator acts on a SNW vertex and results in another vertex, preserving the combinatorial structure. The general form is 
\begin{equation}\label{2nd}
\hat{\mathcal{O}}_t\equiv\int dg_vdg_wd\phi_vd\phi_w O(g_v,g_w;\phi_v,\phi_w)\hat{\varphi}_t^{\dagger}(g_v,\phi_v)\hat{\varphi}_t(g_w,\phi_w),
\end{equation}
where 
\begin{equation}
O(g_v,g_w;\phi_v,\phi_w)\equiv\bra{g_v;\phi_v}\hat{O}\ket{g_w;\phi_w}
\end{equation}
is the matrix element of a first quantized geometric operator for a single tetrahedron, in the one-body Hilbert space.\\

Since a shell state contains three layers, some of the geometric operators will explicitly depend on only one of them. 
For example, the area of the outer boundary of a shell should only have contributions from the outer boundary, while a generic operator not appropriately restricted would add the areas of all surfaces in all the three layers. Restricted area operators would instead only depend on condensate wavefunctions of a subset of the tetrahedra. By including the appropriate $\hat{\sigma}_{ts}$ and $\hat{\sigma}^{\dagger}_{ts}$ in the expression, the area is correspondingly restricted to measure quantities on a specific layer \cite{condensate}:
\begin{equation}\label{total}
\hat{\mathbb{O}}_{u,ts}\equiv\int dh_vdh_w O_{us}(h_v,h_w)\hat{\sigma}_{u,ts}^{\dagger}(h_v)\hat{\sigma}_{u,ts}(h_w),
\end{equation}
where 
\begin{eqnarray}
O_{us}(h,h')&=&\tilde{\varsigma}_{us}^3\prod_{s'\neq s}\tilde{\varsigma}_{us'}^2\nonumber\\
&&\int dg_vdg_wd\phi_vd\phi_w \sigma_{u,s}(h_vg_v,\phi_v)O(g_v,g_w;\phi_v,\phi_w)\overline{\sigma_{u,s}(h_wg_w,\phi_w)}
\end{eqnarray}
by requiring that $\bra{\tau_u}\hat{\mathbb{O}}_{u,ts}\ket{\tau_u}\sim\int dh dh'\bra{0}\hat{\sigma}_{u,ts}(h)\hat{\mathcal{O}}_t\hat{\sigma}^{\dagger}_{u,ts}(h')\ket{0}$. These two values differ by some Dirac delta functions due to connectivity. For instance, the number operator counting the vertices with a specific colour on a layer reads
\begin{equation}\label{numberf}
\hat{n}_{u,ts}\equiv\tilde{\varsigma}_{us}^3\prod_{s'\neq s}\tilde{\varsigma}_{us'}^2\int dh_v  \hat{\sigma}^{\dagger}_{u,ts}(h_v)\hat{\sigma}_{u,ts}(h_v),
\end{equation}
so $\bra{\Gamma_n(\phi_u)}\hat{n}_{u,ts}\ket{\Gamma_n(\phi_u)}=n$, as expected. 

\subsection{Area and volume of a shell state}
For our purposes we focus on area and volume operators only.
Since the states in SNW basis diagonalises these two operators, let us turn to the spin representation to simplify the calculation. In this representation, the basic group creation operators $\hat{\varphi}^{\dagger}_t(g,\phi)$ are
\begin{eqnarray}
&&\hat{\varphi}_t^{\dagger}(g,\phi)=\sum_{j,m,n,l}\hat{\varphi}_{(t)m_1m_2m_3m_4}^{\dagger j_1j_2j_3j_4l}(\phi)
\iota_{n_1n_2n_3n_4}^{j_1j_2j_3j_4l}\prod_{i=1}^{4}\overline{D_{m_in_i}^{j_i}(g_i)},
\end{eqnarray}
which satisfy
\begin{equation}
[\hat{\varphi}_{(t)m_1m_2m_3m_4}^{\quad j_1j_2j_3j_4l}(\phi),\hat{\varphi}_{(t')m_1'm_2'm_3'm_4'}^{\dagger j_1'j_2'j_3'j_4'l'}(\phi')]=\delta_{t,t'}\delta_{j_i,j_i'}\delta_{m_i,m_i'}\delta_{l,l'}\delta^4(\phi-\phi').
\end{equation}
Therefore, the condensate field operator reads
\begin{eqnarray}
\hat{\sigma}_{t}(h_I)&=&\int dg_I d\phi \sum_{j,m,n,o,l_L,l_R}\sigma^{j_1j_2j_3j_4l_Ll_R}(\phi)\iota_{m_1m_2m_3m_4}^{j_1j_2j_3j_4l_L}\iota_{n_1n_2n_3n_4}^{j_1j_2j_3j_4l_R}\nonumber\\
&&\prod_{I=1}^{4}D_{m_Io_I}(h_I)D_{o_In_I}(g_I)\hat{\varphi}_t(g_I,\phi)\nonumber\\
&\equiv&\sum_{j,m,o,l_L}\hat{\tilde{\sigma}}^{\quad j_1j_2j_3j_4l_L}_{(t)o_1o_2o_3o_4}\iota_{m_1m_2m_3m_4}^{j_1j_2j_3j_4l_L}\prod_{I=1}^4D_{m_Io_I}^{j_I}(h_I),
\end{eqnarray}
where the wavefunction assigned to the new operator is
\begin{equation}
\tilde{\sigma}_{(t)o_1o_2o_3o_4}^{j_1j_2j_3j_4l_L}(g_I,\phi)
=\sum_{l_R}\sigma^{j_1j_2j_3j_4l_Ll_R}(\phi)\iota_{n_1n_2n_3n_4}^{j_1j_2j_3j_4l_R}\prod_{I=1}^4D_{o_In_I}^{j_I}(g_I).
\end{equation}
The operator can be extended as
\begin{eqnarray}
\hat{\tilde{\sigma}}_{(u,t)o_1o_2o_3o_4}^{\dagger j_1j_2j_3j_4l_L}
=\sum_{l_R}\int dg_I d\phi\sigma_{u}^{j_1j_2j_3j_4l_Ll_R}(\phi)\iota_{n_1n_2n_3n_4}^{j_1j_2j_3j_4l_R}\prod_{I=1}^4D_{o_In_I}^{j_I}(g_I)\hat{\varphi}^{\dagger}(g_I,\phi),
\end{eqnarray}
which peaks at $\phi_u$.\\

Now let us find the area and the volume expectation of a shell $u$. The area operator counts the area of triangles on the outer boundary whose dual links are labeled by $1$, with definition
\begin{equation}
\hat{\mathbb{A}}_{1}(\phi_u)=\sum_{t=W,B}\hat{\mathbb{A}}_{1,t+}\equiv 
\prod_{s=\{+,0,-\}}\tilde{\varsigma}_{us}^2
\kappa  \sum_{t=B,W}\int dh_{I}^{v}\hat{\sigma}_{r,t+}^{\dagger}(h_{I}^v)\sqrt{E_1^iE_1^j\delta_{ij}}\rhd\hat{\sigma}_{r,t+}(h_I^v),
\end{equation}
where $i=\{1,2,3\}$. Here, $\kappa$ is a multiple of the Planck length square (if adopting the loop quantization scheme, $\kappa=8\pi\gamma \ell_P^2$, with Barbero-Immirzi parameter $\gamma$). Meanwhile, $E_1^i$ is the flux or discrete triad operator, proportional to Lie derivatives on $SU(2)$, whose action is \cite{bhc,condensate,bhlett}
\begin{equation}
E_1^i \rhd f(g_I):= \lim_{\epsilon\to 0}i \frac d{d\epsilon}f(e^{-i\epsilon\tau^i}g_1,...,g_4),
\end{equation}
for a function $f: SU(2)^4\to \mathbb{C}$, where $\tau^{i}$ is the Pauli matrix, while $\tau^{i}\tau^{i}=j(j+1)\mathbbm{1}$ is the $SU(2)$ Casimir operator. The area expectation value reads
\begin{eqnarray}\label{exp2}
&&\bra{\Gamma_n(\phi_u)}\hat{\mathbb{A}}_{1}(\phi_u)\ket{\Gamma_n(\phi_u)}\nonumber\\
&=&\sum_{t=\{B,W\}}\kappa\langle\hat{n}_{u,ts}\rangle\int dg_I^v dh_I^v d\phi_v\sigma_{\phi_u+\delta \phi_u}(h_I^vg_I^v,\phi_v)\sqrt{E_1^iE_1^j\delta_{ij}}\rhd \overline{\sigma_{\phi_u+\delta \phi_u}(h_I^vg_I^v,\phi_v)}\nonumber\\
&\equiv&\sum_{t=\{B,W\}}\langle\hat{n}_{u,t+}\rangle a_{1,u+},
\end{eqnarray}
where $a_{1,us}$ is the expectation value for a single radial link $1$ in the boundary $+$ of shell $u$, peaking at $\phi_u+\delta \phi_u$:
\begin{eqnarray}
a_{1,u+}&=&\kappa \sum_{l_L,j,o} \int dh d\phi  \tilde{\sigma}_{(u+)o_1o_2o_3o_4}^{j_1j_2j_3j_4l_L}(h,\phi)\sqrt{E_1^iE_1^j\delta_{i,j}}\rhd \overline{\tilde{\sigma}_{(u+)o_1o_2o_3o_4}^{j_1j_2j_3j_4l_L}(h,\phi)}\nonumber\\
&\simeq&\kappa \tilde{\varsigma}_{u+}^{-1}\sum_{l_L,j,o,n,n'}\int dh
\frac{d_{j_1}^2d_{j_2}^2d_{j_3}^2d_{j_4}^2}{n(j_1,j_2,j_3,j_4,l_L)}\nonumber\\
&&\iota_{n_1n_2n_3n_4}^{j_1j_2j_3j_4l_L}
\iota_{n_1'n_2'n_3'n_4'}^{j_1j_2j_3j_4l_L}\prod_{I=1}^{4}D_{o_In_I}^{j_I}(h)\overline{D_{o_In_I'}^{j_I}(h)}\sqrt{j_1(j_1+1)}\nonumber\\
&=&\kappa\tilde{\varsigma}(\phi_u+\delta \phi_u)^{-1}\sum_{j}d_{j_1}d_{j_2}d_{j_3}d_{j_4}\sqrt{j_1(j_1+1)}.
\end{eqnarray}
Thus, $\hat{\mathbb{A}}(\phi_u)$ acts on a shell labeled by $u$, and measures the area of a layer whose `radial coordinate' is $\phi_u+\delta \phi_u$. As a result,
\begin{equation}\label{area}
\langle{\mathbb{A}}_{sphere}(\phi_u)\rangle\propto \tilde{\varsigma}^{-1}(\phi_u).
\end{equation}
The volume expectation value can be calculated in a similar way. The volume (density) operator reads
\begin{equation}
\hat{\mathbb{V}}_{u,ts}=\int dh_v dh_w \hat{\sigma}^{\dagger}_{u,ts}(h_v)V(h_v,h_w)\hat{\sigma}_{u,ts}(h_w),
\end{equation}
which results in
\begin{eqnarray}
\langle\hat{\mathbb{V}}_{u,ts}\rangle&=&\langle\hat{n}_{u,ts}\rangle
\int dh dh' dg dg\ d\phi d\phi' \delta(h,h') \sigma_{us}(hg,\phi)V(g,g')\overline{\sigma_{us}(h'g',\phi')}\nonumber\\
&\equiv&\langle\hat{n}_{u,ts}\rangle \mathcal{V}_{(a)ts}(\phi_u).
\end{eqnarray}
The spatial-volume then reads
\begin{eqnarray}
\langle V_3(\phi_u)\rangle&=&\sum_{t=\{B,W\}}\sum_{s=\{+,0,-\}}\langle\hat{n}_{u,ts}\rangle \mathcal{V}_{(a)ts}(\phi)\delta \phi|_{\phi=\phi_u},
\end{eqnarray}
where $\delta\phi$ is the infinitesimal variation in field space \cite{density} and it should not be confused with $\delta\phi_u$, a fixed small real number. Letting $V_{j_1j_2j_3j_4}$ be the eigenvalue of a volume density operator acting on a SNW labeled by $j_1$, $j_2$, $j_3$, and $j_4$, one obtains
\begin{eqnarray}
\mathcal{V}_{(a)us}&=&\sum_{j,j',m,m',l,l'}\int dh dh' d\phi_vd\phi_w \delta_{j,j'}\delta_{m,m'} V_{j_1j_2j_3j_4}
\nonumber\\
&&\tilde{\sigma}_{(u,ts)m_1m_2m_3m_4}^{j_1j_2j_3j_4l}(h,\phi_v))\delta(l,l')\overline{\tilde{\sigma}_{(u,ts)m_1'm_2'm_3'm_4'}^{j_1'j_2'j_3'j_4'l'}(h,\phi_w)}\nonumber\\
&=&\sum_{j,m,l_R,l_R',n,n'}\int dh  d\phi_w \delta(l_R,l_R')\sigma_{us}^{j_1j_2j_3j_4l_R}(\phi_w)
\overline{\sigma_{us}^{j_1j_2j_3j_4l_R'}(\phi_w)}\nonumber\\
&&\iota_{n_1n_2n_3n_4}^{j_1j_2j_3j_4l_R}\iota_{n_1'n_2'n_3'n_4'}^{j_1j_2j_3j_4l_R'}
\prod_{i=1}^{4}D_{m_in_i}^{j_i}(h_i)\overline{D_{m_in_i'}^{j_i}(h_i)}
V_{j_1j_2j_3j_4}\nonumber\\
&\simeq&\tilde{\varsigma}_{us}^{-1}\sum_{j}V_{j_1j_2j_3j_4}d_{j_1}d_{j_2}d_{j_3}d_{j_4}.
\end{eqnarray}
So the total shell-$u$ volume reads
\begin{eqnarray}\label{volume}
\langle V_3(\phi_u)\rangle&\propto&\tilde{\varsigma}(\phi)^{-1}\delta\phi|_{\phi=\phi_u}
+\tilde{\varsigma}(\phi)^{-1}\delta\phi|_{\phi=\phi_u+\delta\phi_u} +\tilde{\varsigma}(\phi)^{-1}\delta\phi|_{\phi=\phi_u-\delta\phi_u}\nonumber\\
&\simeq&3\tilde{\varsigma}(\phi)^{-1}\delta\phi|_{\phi=\phi_u}.
\end{eqnarray}

The next step will be to move to the classical side of the story, consider possible relational descriptions of spherically symmetric continuum geometries in difference matter reference frames, and identify which choice leads to a good match for the observables we computed, under suitable restrictions for $\sigma$,

\subsection{Gluing of shells, spherically symmetric states and their properties}
Before we do so, let us conclude our analysis of the properties of these candidate microstates for spherically symmetric geometries, considering the case in which horizon-like conditions are imposed. We will find that, also in this generalised formulation using matter reference frames, the results of the original construction \cite{bhc} can be reproduced. In particular, we can show that our quantum microstates manifest holographic properties and an area law for their horizon entropy. Since the procedure is in fact also analogous to the one in \cite{bhc}, to which we refer for more details, we limit ourselves to a very brief summary of calculations and results. \\

To form a full spherically symmetric 3-space, one glues the shells. 
The generic complete state can be written as 
\begin{equation}
\ket{\tilde{\Psi}}=\prod_u\ket{\Psi_u},
\end{equation}
with density matrix
\begin{equation}
\hat{\rho}=\ket{\tilde{\Psi}}\bra{\tilde{\Psi}}.
\end{equation}
In order to understand the entanglement structure between shells, we consider a graph $A$ describing the outer boundary of shell at $\phi_1$, and graph $B$ for inner boundary of shell at $\phi_2$. Assume both graphs consist of $n$ vertices so that they could be glued properly. Now the wavefunction for this glued graph is
\begin{eqnarray}
&&\psi(g_I^{A_1},...,g_I^{A_n},g_I^{B_1},...,g_I^{B_n};\phi^{A_1},...\phi^{A_n},\phi^{B_1},...\phi^{B_n})\nonumber\\
&=&\prod_{i=1}^n\tilde{\sigma}_{A_i m_1^im_2^im_3^im_4^i}^{j_1j_2j_3j_4l_R}(g_I^{A_i},\phi^{A_i})
\tilde{\sigma}_{B_i n_1^in_2^in_3^in_4^i}^{j_1j_2j_3j_4l_R}(g_I^{B_i},\phi^{B_i})\\
&&\delta_{m_1^i,-n_1^{t_1^m(i)}}\prod_{J=2}^4\delta_{m_J^i,-m_J^{t_J^m(i)}}\delta_{n_J^i,-n_J^{t_J^n(i)}},
\end{eqnarray} 
where $t_J^m(i)/t_J^n(i)$ denoting the target vertex in graph $A/B$ of edge $J$ departing from vertex $i$, and $\delta$'s are used to keep track of the connectivity of graph $A\cup B$.\\

Integrating away the $B$ part leaves a reduced density matrix for part $A$, where
\begin{eqnarray}\label{reduced matrix}
&&\rho_A^{(n)}(g_I^1,...,g_I^n,{g'}_I^1,...,{g'}_I^n;\phi^1,...\phi^n,{\phi'}^1,...,{\phi'}^n)\nonumber\\
&=&\left(\frac{\tilde{\varsigma}(\phi_1+\delta \phi_u)}{\prod_{I=1}^4d_{j_I}}\right)^n
\prod_{i=1}^n\tilde{\sigma}_{A_i m_1^im_2^im_3^im_4^i}^{j_1j_2j_3j_4l_L}(g_I^{i},\phi^i)\overline{\tilde{\sigma}_{A_i m_1^{'i}m_2^{'i}m_3^{'i}m_4^{'i}}^{j_1j_2j_3j_4l_L}({g'}_I^{i},{\phi'}^{i})}\nonumber\\
&&\delta_{m_1^i,{m'}_1^i}\prod_{J=2}^4\delta_{m_J^i,-m_J^{t_J^m(i)}}\delta_{{m'}_J^i,-{m'}_J^{t_J^{m'}(i)}}.
\end{eqnarray}
This indicates that no information about bulk degrees of freedom can be found from a reduced density matrix, so only the shells nearby contribute to the entanglement entropy \cite{bhc}. \\

For a part $A$ with combinatorial pattern $\alpha$, it has eigenstates
\begin{equation}\label{eistate}
\Psi_{u,s}^n(\Gamma_{alpha})=\Psi_A^{(n)}(n_1,g,\phi)=\left(\frac{\tilde{\varsigma}(\phi_1+\delta \phi_u)}{\prod_{I=1}^4 d_{j_I}}\right)^{\frac n2}\prod_{I=1}^{n}\overline{\tilde{\sigma}_{A_in_1^in_2^in_3^in_4^i}^{j_1j_2j_3j_4l_L}(g_I^i,\phi^i)}\prod_{J=2}^4\delta_{n_J^i,-n_J^{t_J^n(i)}},
\end{equation}
which satisfies
\begin{equation}
\bra{\Psi_A^{(n)}(n_1,g,\phi')}\ket{\Psi_A^{(n)}(n'_1,g,\phi)}=\prod_{i=1}^n\delta_{n_1^i,{n'}_1^i}.
\end{equation}
We find that
\begin{eqnarray}
&&\int \prod_{i=1}^ndg^id\phi^i\rho_A^{(n)}(g_I^1,...,g_I^n,{g'}_I^1,...,{g'}_I^n;\phi^1,...\phi^n,{\phi'}^1,...,{\phi'}^n)\Psi_A^{(n)}(n_1,g,\phi)\nonumber\\
&=&\Psi_A^{(n)}(m_1,g',\phi').
\end{eqnarray}
It is proved that \cite{bhc}
\begin{equation}
\bra{\Psi^n_{u,s}(\Gamma_{\alpha})}\ket{\Psi^n_{u,s}(\Gamma_{\alpha'})}=\delta_{\alpha,\alpha'}\prod_{i=1}^n\delta_{n_1^i,{n'}_{1}^{i}},
\end{equation}
which implies
\begin{equation}
\rho_{u,s}^{(n)}(\Gamma_{\alpha})\Psi_{u,s}^{(n)}(\Gamma_{\alpha'}) =\left\{
\begin{array}{lcl}
\Psi_{u,s}^{(n)}(\Gamma_{\alpha'})\quad   &\alpha=\alpha' \\
0\quad  &\alpha\not=\alpha'
\end{array}.
\right.
\end{equation}
The above computation shows that the scalar fields degrees of freedom do not affect the entanglement structure of the states, and the reduced density matrix can be diagonalized, just like it was possible in their absence. Therefore, the entropy of a shell can be obtained again by counting the number of the states (or graphs) \cite{bhc}. \\

One can then identify a given shell state as a horizon by imposing a condition of maximal entropy, as a proxy for a full local and complete characterization, which is missing in our context. The resulting entropy can be computed, then, under the constraint of fixed total area, and straightforwardly if the combinatorial structures of the three layers are the same. The details can be found in \cite{bhc}. The result is, without further assumption, an area law and a logarithmic correction. An additional semiclassicality condition allows to fix the numerical coefficient of the area law, obtaining
\begin{equation}
S(A_H)\simeq \frac{A_H}{4\ell_P}-\frac{3}{2}\log(\frac{A_H}{\ell_P^2}),
\end{equation}
where $A_H$ is the horizon area, and $\ell_P$ is the Planck length. The logarithm correction is consistent with result in LQG \cite{lqgentropy1,lqgentropy2,lqgentropy3}. 


\section{Recovery of the Schwarzschild Geometry}
\subsection{Klein-Gordon equations}
The actual mean values of both area and volume of our quantum shell states depend on the condensate wavefunction $\sigma$. This remains arbitrary in our construction, so far, also because there is no quantum dynamics providing further information. Absent this, we can try to fix it by matching the computed observables with classical ones. We need therefore a relational description of areas and volumes in a Schwarzschild geometry written in terms of a matter reference frame constituted by four scalar fields. Our goal now is to determine if there exist scalar fields that give a material reference frame in which the Schwarzschild geometry has area and volume observables matching the ones we computed from our quantum microstates. This is of course only a second-best criterion to confirm our interpretation of them as microstates for spherically symmetric geometries, the best being to show that they solve a quantum dynamics reproducing GR (coupled to scalar fields) in a semiclassical approximation, maybe after reduction to a spherically symmetric sector. Still, it is a non-trivial test, and the best we can do at the moment, since solving the GFT quantum dynamics applied to our quantum microstates is out of immediate reach. \\

We want to identify four scalar fields, solving the Klein-Gordon equation, for some choice of potential, in the background of the Schwarzschild geometry. Recall, in fact, that the scalar fields are assumed to provide an (almost) ideal set of clocks and rods, and this means that we are assuming their energy-momentum tensor, hence their backreaction of spacetime geometry, to be negligible. Further, in the same approximation, the corresponding material reference frame can be understood as corresponding to some specific coordinate frame. The Schwarzschild metric:
\begin{equation}\label{schwarz}
ds^2=-\left(1-\frac{2GM}{r}\right)dt^2+\left(1-\frac{2GM}{r}\right)^{-1}dr^2+r^2(d\theta^2+\sin[2](\theta)d\chi^2),
\end{equation}
can be rewritten in the general foliation picked up by the GFT states, that is with some generic time and radial dependence, but homogeneous in the spherical directions:
\begin{equation}\label{radial}
ds^2=-f_1(\tau,u)d\tau^2+f_2(\tau,u)du^2+2f_3(\tau,u)d\tau du+r^2(\tau,u)d\Omega^2,
\end{equation}
where $\tau$, $u$ are arbitrary timelike and spacelike coordinates. By requiring $f_1>0$ and $f_2>0$, the constant-$\tau$ hypersurface is spacelike, and the metric (\ref{radial}) foliates a space into nested spheres, compatible with GFT picture. \\

For the scalar fields dynamics, encoded in the action
\begin{equation}
S_{\phi}=\frac 12 \int \sqrt{-g}\left[(\partial\phi)^2+(m^2+\tilde{\alpha} R)\phi^2\right]
+\int\sqrt{-g}V(\phi)+...,
\end{equation}
with $m$ the mass, $\tilde{\alpha}$ the coupling constant, and $V(\phi)$ a potential term depending on $\phi$, we can only assume the coupling constant $\tilde{\alpha}$ to be zero, i.e. that the scalar field should be `minimally coupled', but we have to reason for further restriction (e.g. that their mass or potential is vanishing, which yields the harmonic gauge: $\nabla_{\mu}\nabla^{\mu}\phi=0$, as it has been possible to do, thanks to the control over the dynamics, in the cosmological setting \cite{luca}). In addition, the scalar field solutions should also be monotonic in temporal or spatial directions, to be used as a clock or rod. Thus the scalar fields must satisfy the following equations:
\begin{equation}
\nabla_{\mu}\nabla^{\mu}\phi_i-m_i^2\phi_i-V_i'(\phi_i)=0,
\end{equation}
\begin{equation}
R_{\mu\nu}-\frac{1}{2}g_{\mu\nu}R=\frac{1}{2}\sum_{i=0}^3\left[\partial_{\mu}\phi_i\partial_{\nu}\phi_i
-\frac{1}{2}g_{\mu\nu}\left((\partial\phi_i)^2+m_i^2\phi_i^2+2V_i(\phi_i)\right)\right]\simeq 0.
\end{equation}

After these equations are solved analytically, the spatial part of the metric \eqref{radial} can be reformulated in a `relational' manner with respect to the scalar fields themselves as
\begin{eqnarray}\label{relationalmetric}
dh^2_{relational}&=&\frac{f_2[\tau(\phi_0),u(\phi_1)]}{\phi_1'(u)^2}d\phi_1^2
+f_4[\tau(\phi_0),u(\phi_1)]r(\phi_0,\phi_1)^2d\Omega^2(\phi_2,\phi_3).
\end{eqnarray}
Using this form of the metric, we can then compute area and volume observables to be matched with the ones computed in our GFT quantum microstates.\\

According to the relational metric (\ref{relationalmetric}), at time $\phi_0=\phi_{\tau}$, the area and the volume of a sphere with $\phi_1=\phi_u$ are
\begin{eqnarray}
&&A_{rel}(\phi_u)=4\pi r^2(\phi_\tau,\phi_u),\\
&&V_{3rel}(\phi_u)=4\pi r^2(\phi_\tau,\phi_1)\frac{\sqrt{f_2[\phi_{\tau},\phi_1]}}{\phi_1'(u)}\delta\phi_1|_{\phi_1=\phi_u}.
\end{eqnarray}
By requiring that
\begin{equation}\label{condition1}
A_{rel}(\phi_u)\propto \langle\mathbb{A}_{sphere}(\phi_u)\rangle_{GFT},
\end{equation}
and 
\begin{equation}\label{condition2}
V_{3rel}(\phi_u)\propto \langle V_3(\phi_u)\rangle_{GFT},
\end{equation}
we can then deduce the form of the wavefunctions that should be assigned to the GFT building blocks.\\

Further conditions are needed, in order to resolve remaining ambiguities and determine uniquely the scalar fields. 
In addition, it must be noted that (\ref{area}) and (\ref{volume}) are invariant under a scalar field redefinition $\tilde{\phi}(\phi)$. 
\\

It is important to note that there is no need to find all scalar fields: $\phi_0$ and $\phi_1$ are sufficient to determine the relational geometry, due to spherical symmetry. 
Also, an exact solution for $\phi_0$ is necessary, even though the geometry is expected to be static, because one needs a clock to define a equal-time hypersurface, to which our GFT states are associated. 
\\

Condition (\ref{condition1}) demands that the usual radial coordinate $r$ should have no $\phi_0$-dependence and thus no $\tau$-dependence. So without losing any generality, one can choose $u=r$. The foliated expression (\ref{radial}) becomes
\begin{equation}\label{metric}
ds^2=-f_1(r)d\tau^2+f_2(r)dr^2+2f_3(r)d\tau dr+r^2d\Omega^2.
\end{equation}
Assume then that $\phi_0(\tau)$ is a scalar field with a potential $V(\phi_0)$ and a mass term $m_0$. Its Klein-Gordon equation reads
\begin{eqnarray}
&&\nabla_{\mu}\nabla^{\mu}\phi_0-V'(\phi_0)-m_0^2\phi_0(\tau)\nonumber\\
&=&-\frac{f_2(r)}{f_1(r) f_2(r)+f_3(r)^2}\phi_0''(\tau)-V'(\phi_0)
-m_0^2\phi_0(\tau)\nonumber\\
&&-\frac{r f_2(r) f_3(r) f_1'(r)+f_1(r) \left(r  f_3(r)f_2'(r)-2 f_2(r) \left(r  f_3'(r)+2 f_3(r)\right)\right)-4  f_3(r)^3}{2 r \left(f_1(r)f_2(r)+ f_3(r)^2\right)^2}\phi_0'(\tau)\nonumber\\
&=&0.
\end{eqnarray} 
Since $f_2(r)>0$, the first term is non-zero unless $\phi_0''(\tau)=0$, but this is impossible if $\phi_0$ is monotonic; so this term cannot vanish. As $\phi_0$ and $V(\phi_0)$ have no $r$-dependence while other terms contain $r$, in order that the equation is consistent, both potential and mass should vanish. For the same reason, the remaining terms containing $r$ should be canceled. A solvable option is that the last term equals zero for any $r\in\mathbb{R}^+$. The equation
\begin{eqnarray}
r f_2(r) f_3(r) f_1'(r)+f_1(r) \left(r  f_3(r)f_2'(r)-2 f_2(r) \left(r  f_3'(r)+2 f_3(r)\right)\right)-4  f_3(r)^3=0
\end{eqnarray}
has three solutions:
\begin{eqnarray}
&&f_3(r)=0,\\
&&f_3(r)=\frac{\sqrt{f_1(r)f_2(r)} }{\sqrt{-1+C_1 r^4}},\\
&&f_3(r)=-\frac{\sqrt{f_1(r)f_2(r)} }{\sqrt{-1+C_1 r^4}},
\end{eqnarray}
where $C_1$ is a constant. Both $\phi_0$ and $\phi_1$ must be real, so the last two choices, yielding a solution $\phi_1$ with non-vanishing imaginary part, are therefore ruled out. When $f_3=0$, the Schwarzschild metric in usual spherical coordinates (\ref{schwarz}) is recovered, where $\tau=t$. The Klein-Gordon equation of $\phi_0(t)$ reads
\begin{equation}
\phi_0''(t)=0,
\end{equation}
whose solution is
\begin{equation}
\phi_0(t)=\beta_0 t,
\end{equation}
where $\beta_0$ is a small constant such that $\beta_0^2\ll 1$ so the energy-momentum tensor $T_{tt}\simeq 0$.\\

Then let us move on to find $\phi_1$. Conditions (\ref{condition1}) and (\ref{condition2}) together imply
\begin{equation}
\frac{d\phi_1(r)}{dr}=\frac{\beta_1}{\sqrt{f(r)}},
\end{equation}
where
\begin{equation}
f(r)=1-\frac{2GM}{r},
\end{equation}
with $\beta_1$ another arbitrary constant.
It is immediate to obtain the solution
\begin{equation}\label{s1}
\phi_1(r)=\beta_1 r \sqrt{1-\frac{2 GM}{r}}+2\beta_1 GM \text{arctanh}\left(\sqrt{1-\frac{2 M}{r}}\right),
\end{equation}
for positive mass $M$ with $r>2GM$. Absorbing the mass term in potential term $V(\phi_1)$, one has
\begin{equation}\label{kg}
\nabla_{\mu}\nabla^{\mu}\phi-V'(\phi)=0.
\end{equation}
Substituting the solution into this equation gives the potential
\begin{equation}
\beta_1\frac{2 r-3 GM}{r^2\sqrt{f(r)}}=V'(\phi_1).
\end{equation}

The inverse $r(\phi_1)$ tells how area and volume of a sphere/shell depend on $\phi_1$. Although the inverse cannot be expressed analytically at the moment, we can find it by taking different limits. First is the asymptotic infinity $r\to\infty$. In this region, $V'(\phi_1)\simeq0$, and the simplest choice is that $V(\phi_1)=0$. Then for the rod, we have $\phi_1'(r)\simeq 0$, and
\begin{equation}
\phi_1(r)\simeq\beta_1r.
\end{equation}
Therefore, $\phi_1\to\infty$ corresponds the asymptotic infinity. While in the near-horizon regime, one finds $f(r)\simeq r/2M-1$, and
\begin{equation}
\frac{d\phi_1(r)}{dr}\simeq\frac{\beta_1}{\sqrt{\frac{r}{2GM}-1}}.
\end{equation}
Hence,
\begin{equation}
\phi_1(r)\simeq 4\beta_1 GM \sqrt{\frac{r}{2M}-1}.
\end{equation}
At the event horizon one obtains $\phi_1\simeq0$. 
The inverse reads
\begin{equation}
r= \frac{\phi_1^2}{8 \beta_1^2 GM}+2 GM.
\end{equation}
Consequently,
\begin{equation}
V'(\phi_1)\simeq\frac{\beta_1^2}{\phi_1},
\end{equation}
and
\begin{equation}
V(\phi_1)\simeq \beta_1^2\ln(\phi_1).
\end{equation}

It can be checked that $\phi_0$ and $\phi_1$ are a good pair of clock and rod. As $\phi_0=\beta_0 t$, there is a one-to-one correspondence between $\phi_0$ and $t$. For $\phi_1$, $\phi_1'(r)\propto 1/\sqrt{f(r)}$, and it is always true that $f(r)>0$. So $\phi_1$ is a monotonic function of $r$.\\

To conclude, starting from our kinematical quantum microstates, we were able to identify the precise form of the  continuum scalar fields that give a relational Schwarzschild geometry with a compatible structure. Now we move to the matching of area and volume observables.

\subsection{Matching of observables and conditions on GFT wavefunction}
With these solutions, it is straightforward to compute relational observables and match them with the expectation values of the corresponding quantum observables in our microstates. We do so considering the geometry at asymptotic infinity and at the (near-)horizon, since our clocks and rods are only valid outside the horizon\footnote{Because $\phi(r)$ ends at $r=2GM$, the rod cannot enter the horizon (calculations and observations on rods cannot be made inside the event horizon). }.\\

The relational spatial metric reads
\begin{equation}
dh^2_{\text{rel}}=\frac{1}{\beta_1^2}d\phi_1^2+r^2(\phi_1)d\Omega^2.
\end{equation}
At asymptotic infinity, $\phi_1\to\infty$, a sphere at $\phi_1=\phi_u$ has the area and the $3$-volume 
\begin{eqnarray}
&&A_{rel}(\phi_u)=4\pi \left(\frac{\phi_u}{\beta_1}\right)^2,\\
&&V_{3rel}(\phi_u)=\frac{4\pi}{\beta_1}\left(\frac{\phi_1}{\beta_1}\right)^2\delta\phi_1|_{\phi_1=\phi_u}.
\end{eqnarray}
While in near-horizon regime when $M>0$,
\begin{eqnarray}
&&A_{rel}(\phi_u)=4\pi \left(\frac{\phi_u^2}{8 \beta_1^2 GM}+2 GM\right)^2,\\
&&V_{3rel}(\phi_u)=\frac{4\pi}{\beta_1}\left(\frac{\phi_1^2}{8 \beta_1^2 GM}+2 GM\right)^2\delta\phi_1
|_{\phi_1=\phi_u}.
\end{eqnarray}
requiring good matching with the results of the quantum calculation, we determine the required form of the GFT condensate wavefunction (in its dependence on the scalar field variables, the only aspect left to be determined) to be:
\begin{align}
\begin{split}
\text{Schwarzschild} \left \{
\begin{array}{ll}
\text{Asymptotic infinity}(\phi\to \infty)&\tilde{\varsigma}(\phi)^{-1}\propto \phi^2\\
\text{Near-horizon}(\phi\sim 0)&\tilde{\varsigma}(\phi)^{-1}\propto 
(\phi^2+16\beta_1^2(GM)^2)^2\\
\text{Near-singularity}&\text{Not available}
\end{array}
\right.
\end{split}
.
\end{align}

One final comment is that the GFT states are constructed to be regular, i.e. to have non-vanishing areas for any shell surface. 
For the situation discussed above, this is because we cannot know, measure, or observe whether the rod can extend inside the horizon. However, even if we consider a negative-mass black hole, where the same procedure leads to $\phi\to 0$ and $\langle A_{rel}(\phi)\propto\phi^{4/3}\rangle$ near singularity, no singular state exists either. One always measure the area of the outer boundary of a shell, so the smallest value of scalar field on this layer is $2\delta \phi_u$ when the inner boundary labeled by $\phi=0$, still resulting in a non-zero area. In this sense, there is no shell state corresponding to the curvature singularity $r=0$.\\

\section{Conclusion and Outlooks}
In this paper we have constructed and studied kinematical quantum gravity condensate states for quantum geometry coupled to scalar fields, that represent candidate microstates for spherically symmetric 3-geometries expressed in a matter reference frame. The construction extends earlier ones \cite{bhc} where additional parameters on the quantum microstates, rather than physical degrees of freedom, were used to guide the geometric interpretation, as well as others \cite{steffen} in which much simpler GFT states have been considered for spherically symmetric geometries. We have shown that, besides having interesting holographic properties and an area law for their surface entropy (when this is maximized as horizons are expected to do), our quantum microstates reproduce relevant geometric observables of a continuum Schwarzschild geometry, written in a specific matter reference frame, which we also determined. \\

These results support the geometric interpretation of these quantum microstates, which in turn can now be used for many further analyses with greater confidence and a better geometric intuition. They also provide a concrete example of what we believe are general tasks to be solved for matching fundamental quantum microstates  with effective semiclassical geometries in any fully background independent quantum gravity formalism, via relational, diffeo-invariant observables. \\

A similar study of spherically symmetric spaces via GFT condensates coupled with scalar fields has been conducted in \cite{steffen}, where the constraints on wavefunction and the solutions for scalar fields were found by matching relational observables as well. 
While the methods are similar, our analysis is made more complicated but also, we feel, more interesting, by the more involved structure of our candidate quantum microstates. Indeed, whilst the previous work considers the disconnected tetrahedra (no quantum correlations among spin network vertices are included), the GFT states we considered have intricate connectivity which allows to control the topology of the superposed simplicial complexes dual to the spin network states, in turn a key element for the interpretation as spherically symmetric geometries. Next, the use of coherent peaking state results in controllable quantum fluctuations for large number of GFT quanta (as appropriate in a continuum approximation. Moreover, our result could match both areas and volume with those in the continuum theory, where only the volume was considered in \cite{steffen}. 
Another difference is that the scalar fields in our model do not necessarily satisfy the harmonic gauge\textcolor{red}{,} as well as potential terms are included; in fact, as it turns out they must be included to ensure the expected geometric matching. \\

Still, our analysis has been limited in several aspects, which call for further work. First, our matter frame, in particular, the chosen rod scalar field, cannot extend into the horizon (this was also the case in earlier work \cite{steffen}). In order to obtain real solutions of Klein-Gordon equations, one can only find the scalar fields monotonic in the usual spherical coordinates $\{t,r,\theta,\chi\}$ separately. However, when crossing the horizon, $r$ becomes a temporal coordinate, so $\phi_1(r)$ is no longer a rod. Similarly, the GFT states we constructed cannot describe both interior and exterior Schwarzschild regions either, as it can be seen by looking at their dependence on the scalar field we use as a rod for the radial direction.  The construction is more likely to reach inside the horizon if the scalar fields behave like Kruskal-Szekeres coordinates. A corresponding construction for the GFT condensate states should be sought. Our analysis was also limited to matching simple geometric observables, while we did not match curvature observables, which are of course of great importance when investigating black hole geometries. This is mostly due to the poor control we currently have on curvature operators in GFT (and, in fact, in the related canonical LQG or spin foam models), and one more aspect to improve upon.\\

Obviously, our results, while promising, do not guarantee that the quantum microstates we have studied are the correct ones, even at the mere kinematical level, to describe spherically symmetric quantum geometries and black holes. Different constructions should definitely be explored (in particular, relaxing the condition of microscopic homogeneity for the shells, and looking for a coarse grained encoding of the same property).
Finally, another, more important limitation is that, as already emphasized, the quantum dynamics has not been included in the analysis. This limitation should be necessarily overcome to extract relevant physical consequences of the quantum microstates (and of specific GFT models). Only after some of these limitations are overcome, we can be confident that we achieved a solid, compelling characterization of the microstructure of spherically symmetric geometries in full (GFT and LQG) quantum gravity. 
We will have then a good basis for tacking the outstanding issue in quantum black holes physics. In particular, we could extend to semiclassical black holes the study of perturbations and effective field theory on a quantum gravity-generated spacetime, that was carried out in the GFT cosmology context \cite{gft4, gft6}, remaining indeed within the fundamental quantum gravity formalism, thus having control over limits of validity and possible improvements of our approximations. Also, could study transition mechanisms between our proposed quantum microstates of geometry, studying from this microscopic perspective the evolution and evaporation mechanisms for quantum black holes, aiming at a fundamental quantum gravity resolution of the information loss puzzle appearing in the semiclassical treatment.
Still, despite these limitations, we feel that we have taken one more step in the right direction.


\appendix
\section{Negative mass}
In this appendix, we report the analysis of scalar field frames for negative mass Schwarzschild geometries.\\

The relational description for the singularity is only valid for $M<0$ with a naked singularity \cite{steffen}. When $M$ is negative, one has
\begin{equation}\label{s2}
\phi_1(r)=\sqrt{r(r-2GM)}+2M\text{arcsinh}\left(\sqrt{-\frac{r}{2GM}}\right).
\end{equation}
The Klein-Gordon equation (\ref{kg}) remains unchanged, and the derivatives of solution (\ref{s1}) and (\ref{s2}) with respect to $r$ are the same. The expression (\ref{s1}) is modified to (\ref{s2}) because when $x>1$, $\text{arctanh}(x)$ as a real-valued function is ill-defined.
When $r\to0$, $f(r)\simeq -2GM/r$. As a result, the $\phi_1(r)$ reads
\begin{equation}
\phi_1(r)\simeq\frac{\sqrt{2} \beta_1 r}{3 \sqrt{-GM/r}}.
\end{equation}
Thus, $\phi_1\simeq0$ corresponds to a naked singularity with
\begin{equation}
r=\left(-\frac{9 GM \phi_1^2}{\sqrt{2} \beta_1^{2}}\right)^{\frac{1}{3}}.
\end{equation}
For the potential, one obtains
\begin{equation}
V'(\phi_1)\simeq -\frac{\sqrt{2}\beta_1^2}{\phi_1},
\end{equation}
and 
\begin{equation}
V(\phi_1)\simeq \sqrt{2}\beta_1^2\ln(\frac{1}{\phi_1}).
\end{equation}
In order that the energy-momentum tensor of $\phi_1$ is negligible, $|\beta_1|$ is a very small constant too. Thus, the potential $V(\phi_1)$ at order $\beta_1^2$ can be dropped.\\

Repeating the same procedure, one obtains the near-singularity behaviour when $M<0$, where
\begin{eqnarray}
&&A_{rel}(\phi_u)=4\pi \left(\frac{9 GM \phi_u^2}{\sqrt{2} \beta_1^{2}}\right)^{\frac{2}{3}},\\
&&V_{3rel}(\phi_u)=\frac{4\pi}{\beta_1}\left(\frac{9 GM \phi_1^2}{\sqrt{2} \beta_1^{2}}\right)^{\frac{2}{3}}\delta\phi_1|_{\phi_1=\phi_u}.
\end{eqnarray}
Together with the positive-mass case, the wavefunctions are
\begin{align}
\begin{split}
\text{Schwarzschild}\left \{
\begin{array}{ll}
M>0 &\Big\{ \begin{array}{ll}
\text{Asymptotic infinity}(\phi\to \infty)&\tilde{\varsigma}(\phi)^{-1}\propto \phi^2\\
\text{Near-horizon}(\phi\sim 0)&\tilde{\varsigma}(\phi)^{-1}\propto 
(\phi^2+16\beta_1^2(GM)^2)^2\\
\text{Near-singularity}&\text{Not available}
\end{array}\\
M<0 & \Big\{ \begin{array}{ll}
\text{Asymptotic infinity}(\phi\to \infty)&\tilde{\varsigma}(\phi)^{-1}\propto \phi^2\\
\text{Near-horizon}&\text{Non-existent}\\
\text{Near-singularity}(\phi\sim 0)&\tilde{\varsigma}(\phi)^{-1}\propto 
\phi^{\frac{4}{3}}
\end{array}
\\
\end{array}
\right.
\end{split}
.
\end{align}
 A negative-mass black hole has a naked singularity, so no horizon exists in this case. Hence, no wavefunction corresponds to near-horizon regime.
\acknowledgments
DO acknowledges financial support from the Deutsche Forschung Gemeinschaft (DFG).


\bibliographystyle{JHEP}
\bibliography{GFTBH}

\providecommand{\href}[2]{#2}\begingroup\raggedright\begin{thebibliography}{10}

\bibitem{bekenstein}
J.D.~Bekenstein, \emph{{Black holes and the second law}},
  \href{https://doi.org/10.1007/BF02757029}{\emph{Lett. Nuovo Cim.} {\bfseries
  4} (1972) 737}.

\bibitem{bekenstein2}
J.D.~Bekenstein, \emph{{Black holes and entropy}},
  \href{https://doi.org/10.1103/PhysRevD.7.2333}{\emph{Phys. Rev. D} {\bfseries
  7} (1973) 2333}.

\bibitem{hawking}
S.W.~Hawking, \emph{{Black hole explosions}},
  \href{https://doi.org/10.1038/248030a0}{\emph{Nature} {\bfseries 248} (1974)
  30}.

\bibitem{hawking2}
S.W.~Hawking, \emph{{Particle Creation by Black Holes}},
  \href{https://doi.org/10.1007/BF02345020}{\emph{Commun. Math. Phys.}
  {\bfseries 43} (1975) 199}.

\bibitem{laws}
J.M.~Bardeen, B.~Carter and S.W.~Hawking, \emph{{The Four laws of black hole
  mechanics}}, \href{https://doi.org/10.1007/BF01645742}{\emph{Commun. Math.
  Phys.} {\bfseries 31} (1973) 161}.

\bibitem{condense}
S.~Gielen, D.~Oriti and L.~Sindoni, \emph{{Homogeneous cosmologies as group
  field theory condensates}},
  \href{https://doi.org/10.1007/JHEP06(2014)013}{\emph{JHEP} {\bfseries 06}
  (2014) 013} [\href{https://arxiv.org/abs/1311.1238}{{\ttfamily 1311.1238}}].

\bibitem{luca}
L.~Marchetti and D.~Oriti, \emph{{Effective relational cosmological dynamics
  from Quantum Gravity}},
  \href{https://doi.org/10.1007/JHEP05(2021)025}{\emph{JHEP} {\bfseries 05}
  (2021) 025} [\href{https://arxiv.org/abs/2008.02774}{{\ttfamily
  2008.02774}}].

\bibitem{Seiberg:2006wf}
N.~Seiberg, \emph{{Emergent spacetime}},  in \emph{{23rd Solvay Conference in
  Physics: The Quantum Structure of Space and Time}}, pp.~163--178, 1, 2006,
  \href{https://doi.org/10.1142/9789812706768_0005}{DOI}
  [\href{https://arxiv.org/abs/hep-th/0601234}{{\ttfamily hep-th/0601234}}].

\bibitem{Padmanabhan:2014jta}
T.~Padmanabhan, \emph{{Emergent Gravity Paradigm: Recent Progress}},
  \href{https://doi.org/10.1142/S0217732315400076}{\emph{Mod. Phys. Lett. A}
  {\bfseries 30} (2015) 1540007}
  [\href{https://arxiv.org/abs/1410.6285}{{\ttfamily 1410.6285}}].

\bibitem{Oriti:2013jga}
D.~Oriti, \emph{{Disappearance and emergence of space and time in quantum
  gravity}}, \href{https://doi.org/10.1016/j.shpsb.2013.10.006}{\emph{Stud.
  Hist. Phil. Sci. B} {\bfseries 46} (2014) 186}
  [\href{https://arxiv.org/abs/1302.2849}{{\ttfamily 1302.2849}}].

\bibitem{Oriti:2018dsg}
D.~Oriti, \emph{{Levels of spacetime emergence in quantum gravity}},
  \href{https://arxiv.org/abs/1807.04875}{{\ttfamily 1807.04875}}.

\bibitem{partial1}
C.~Rovelli, \emph{{Partial observables}},
  \href{https://doi.org/10.1103/PhysRevD.65.124013}{\emph{Phys. Rev. D}
  {\bfseries 65} (2002) 124013}
  [\href{https://arxiv.org/abs/gr-qc/0110035}{{\ttfamily gr-qc/0110035}}].

\bibitem{partial2}
C.~Rovelli, \emph{{What Is Observable in Classical and Quantum Gravity?}},
  \href{https://doi.org/10.1088/0264-9381/8/2/011}{\emph{Class. Quant. Grav.}
  {\bfseries 8} (1991) 297}.

\bibitem{Hoehn:2019fsy}
P.A.~Hoehn, A.R.H.~Smith and M.P.E.~Lock, \emph{{Trinity of relational quantum
  dynamics}}, \href{https://doi.org/10.1103/PhysRevD.104.066001}{\emph{Phys.
  Rev. D} {\bfseries 104} (2021) 066001}
  [\href{https://arxiv.org/abs/1912.00033}{{\ttfamily 1912.00033}}].

\bibitem{bhlett}
D.~Oriti, D.~Pranzetti and L.~Sindoni, \emph{{Horizon entropy from quantum
  gravity condensates}},
  \href{https://doi.org/10.1103/PhysRevLett.116.211301}{\emph{Phys. Rev. Lett.}
  {\bfseries 116} (2016) 211301}
  [\href{https://arxiv.org/abs/1510.06991}{{\ttfamily 1510.06991}}].

\bibitem{bhc}
D.~Oriti, D.~Pranzetti and L.~Sindoni, \emph{{Black Holes as Quantum Gravity
  Condensates}}, \href{https://doi.org/10.1103/PhysRevD.97.066017}{\emph{Phys.
  Rev. D} {\bfseries 97} (2018) 066017}
  [\href{https://arxiv.org/abs/1801.01479}{{\ttfamily 1801.01479}}].

\bibitem{gft1}
D.~Oriti, \emph{{The microscopic dynamics of quantum space as a group field
  theory}},  in \emph{{Foundations of Space and Time: Reflections on Quantum
  Gravity}}, pp.~257--320, 10, 2011
  [\href{https://arxiv.org/abs/1110.5606}{{\ttfamily 1110.5606}}].

\bibitem{gft2}
T.~Krajewski, \emph{{Group field theories}},
  \href{https://doi.org/10.22323/1.140.0005}{\emph{PoS} {\bfseries QGQGS2011}
  (2011) 005} [\href{https://arxiv.org/abs/1210.6257}{{\ttfamily 1210.6257}}].

\bibitem{gft3}
D.~Oriti, \emph{{Group Field Theory and Loop Quantum Gravity}},  in \emph{{Loop
  Quantum Gravity}: {The First 30 Years}}, A.~Ashtekar and J.~Pullin, eds.,
  pp.~125--151, WSP (2017),
  \href{https://doi.org/10.1142/9789813220003_0005}{DOI}.

\bibitem{gft4}
S.~Gielen and L.~Sindoni, \emph{{Quantum Cosmology from Group Field Theory
  Condensates: a Review}},
  \href{https://doi.org/10.3842/SIGMA.2016.082}{\emph{SIGMA} {\bfseries 12}
  (2016) 082} [\href{https://arxiv.org/abs/1602.08104}{{\ttfamily
  1602.08104}}].

\bibitem{gft5}
S.~Carrozza, \emph{{Flowing in Group Field Theory Space: a Review}},
  \href{https://doi.org/10.3842/SIGMA.2016.070}{\emph{SIGMA} {\bfseries 12}
  (2016) 070} [\href{https://arxiv.org/abs/1603.01902}{{\ttfamily
  1603.01902}}].

\bibitem{gft6}
D.~Oriti, \emph{{The universe as a quantum gravity condensate}},
  \href{https://doi.org/10.1016/j.crhy.2017.02.003}{\emph{Comptes Rendus
  Physique} {\bfseries 18} (2017) 235}
  [\href{https://arxiv.org/abs/1612.09521}{{\ttfamily 1612.09521}}].

\bibitem{Gurau:2011xp}
R.~Gurau and J.P.~Ryan, \emph{{Colored Tensor Models - a review}},
  \href{https://doi.org/10.3842/SIGMA.2012.020}{\emph{SIGMA} {\bfseries 8}
  (2012) 020} [\href{https://arxiv.org/abs/1109.4812}{{\ttfamily 1109.4812}}].

\bibitem{Rivasseau:2016wvy}
V.~Rivasseau, \emph{{The Tensor Track, IV}},
  \href{https://doi.org/10.22323/1.263.0106}{\emph{PoS} {\bfseries CORFU2015}
  (2016) 106} [\href{https://arxiv.org/abs/1604.07860}{{\ttfamily
  1604.07860}}].

\bibitem{lqgintro1}
C.~Rovelli, \emph{{Loop quantum gravity}},
  \href{https://doi.org/10.12942/lrr-1998-1}{\emph{Living Rev. Rel.} {\bfseries
  1} (1998) 1} [\href{https://arxiv.org/abs/gr-qc/9710008}{{\ttfamily
  gr-qc/9710008}}].

\bibitem{lqgintro2}
T.~Thiemann, \emph{{Lectures on loop quantum gravity}},
  \href{https://doi.org/10.1007/978-3-540-45230-0_3}{\emph{Lect. Notes Phys.}
  {\bfseries 631} (2003) 41}
  [\href{https://arxiv.org/abs/gr-qc/0210094}{{\ttfamily gr-qc/0210094}}].

\bibitem{lqgintro3}
A.~Ashtekar and J.~Lewandowski, \emph{{Background independent quantum gravity:
  A Status report}},
  \href{https://doi.org/10.1088/0264-9381/21/15/R01}{\emph{Class. Quant. Grav.}
  {\bfseries 21} (2004) R53}
  [\href{https://arxiv.org/abs/gr-qc/0404018}{{\ttfamily gr-qc/0404018}}].

\bibitem{lqgintro4}
C.~Rovelli, \emph{Quantum Gravity}, Cambridge Monographs on Mathematical
  Physics, Cambridge University Press (2004),
  \href{https://doi.org/10.1017/CBO9780511755804}{10.1017/CBO9780511755804}.

\bibitem{lqgintro5}
K.~Giesel and H.~Sahlmann, \emph{{From Classical To Quantum Gravity:
  Introduction to Loop Quantum Gravity}},
  \href{https://doi.org/10.22323/1.140.0002}{\emph{PoS} {\bfseries QGQGS2011}
  (2011) 002} [\href{https://arxiv.org/abs/1203.2733}{{\ttfamily 1203.2733}}].

\bibitem{lqgintro6}
N.~Bodendorfer, \emph{{An elementary introduction to loop quantum gravity}},
  \href{https://arxiv.org/abs/1607.05129}{{\ttfamily 1607.05129}}.

\bibitem{2nd}
D.~Oriti, \emph{{Group field theory as the 2nd quantization of Loop Quantum
  Gravity}}, \href{https://doi.org/10.1088/0264-9381/33/8/085005}{\emph{Class.
  Quant. Grav.} {\bfseries 33} (2016) 085005}
  [\href{https://arxiv.org/abs/1310.7786}{{\ttfamily 1310.7786}}].

\bibitem{sf}
A.~Perez, \emph{{The Spin Foam Approach to Quantum Gravity}},
  \href{https://doi.org/10.12942/lrr-2013-3}{\emph{Living Rev. Rel.} {\bfseries
  16} (2013) 3} [\href{https://arxiv.org/abs/1205.2019}{{\ttfamily
  1205.2019}}].

\bibitem{Colafranceschi:2020ern}
E.~Colafranceschi and D.~Oriti, \emph{{Quantum gravity states, entanglement
  graphs and second-quantized tensor networks}},
  \href{https://doi.org/10.1007/JHEP07(2021)052}{\emph{JHEP} {\bfseries 07}
  (2021) 052} [\href{https://arxiv.org/abs/2012.12622}{{\ttfamily
  2012.12622}}].

\bibitem{Colafranceschi:2021acz}
E.~Colafranceschi, G.~Chirco and D.~Oriti, \emph{{Holographic maps from quantum
  gravity states as tensor networks}},
  \href{https://doi.org/10.1103/PhysRevD.105.066005}{\emph{Phys. Rev. D}
  {\bfseries 105} (2022) 066005}
  [\href{https://arxiv.org/abs/2105.06454}{{\ttfamily 2105.06454}}].

\bibitem{Chirco:2021chk}
G.~Chirco, E.~Colafranceschi and D.~Oriti, \emph{{Bulk area law for boundary
  entanglement in spin network states: Entropy corrections and horizon-like
  regions from volume correlations}},
  \href{https://doi.org/10.1103/PhysRevD.105.046018}{\emph{Phys. Rev. D}
  {\bfseries 105} (2022) 046018}
  [\href{https://arxiv.org/abs/2110.15166}{{\ttfamily 2110.15166}}].

\bibitem{condensate}
D.~Oriti, D.~Pranzetti, J.P.~Ryan and L.~Sindoni, \emph{{Generalized quantum
  gravity condensates for homogeneous geometries and cosmology}},
  \href{https://doi.org/10.1088/0264-9381/32/23/235016}{\emph{Class. Quant.
  Grav.} {\bfseries 32} (2015) 235016}
  [\href{https://arxiv.org/abs/1501.00936}{{\ttfamily 1501.00936}}].

\bibitem{con1}
S.~Gielen, D.~Oriti and L.~Sindoni, \emph{{Cosmology from Group Field Theory
  Formalism for Quantum Gravity}},
  \href{https://doi.org/10.1103/PhysRevLett.111.031301}{\emph{Phys. Rev. Lett.}
  {\bfseries 111} (2013) 031301}
  [\href{https://arxiv.org/abs/1303.3576}{{\ttfamily 1303.3576}}].

\bibitem{effective}
D.~Oriti, L.~Sindoni and E.~Wilson-Ewing, \emph{{Emergent Friedmann dynamics
  with a quantum bounce from quantum gravity condensates}},
  \href{https://doi.org/10.1088/0264-9381/33/22/224001}{\emph{Class. Quant.
  Grav.} {\bfseries 33} (2016) 224001}
  [\href{https://arxiv.org/abs/1602.05881}{{\ttfamily 1602.05881}}].

\bibitem{luca2}
L.~Marchetti and D.~Oriti, \emph{{Quantum Fluctuations in the Effective
  Relational GFT Cosmology}},
  \href{https://doi.org/10.3389/fspas.2021.683649}{\emph{Front. Astron. Space
  Sci.} {\bfseries 0} (2021) 110}
  [\href{https://arxiv.org/abs/2010.09700}{{\ttfamily 2010.09700}}].

\bibitem{gplqg}
P.~Martin-Dussaud, \emph{{A Primer of Group Theory for Loop Quantum Gravity and
  Spin-foams}}, \href{https://doi.org/10.1007/s10714-019-2583-5}{\emph{Gen.
  Rel. Grav.} {\bfseries 51} (2019) 110}
  [\href{https://arxiv.org/abs/1902.08439}{{\ttfamily 1902.08439}}].

\bibitem{density}
S.~Gielen and D.~Oriti, \emph{{Cosmological perturbations from full quantum
  gravity}}, \href{https://doi.org/10.1103/PhysRevD.98.106019}{\emph{Phys. Rev.
  D} {\bfseries 98} (2018) 106019}
  [\href{https://arxiv.org/abs/1709.01095}{{\ttfamily 1709.01095}}].

\bibitem{lqgentropy1}
R.K.~Kaul and P.~Majumdar, \emph{{Logarithmic correction to the
  Bekenstein-Hawking entropy}},
  \href{https://doi.org/10.1103/PhysRevLett.84.5255}{\emph{Phys. Rev. Lett.}
  {\bfseries 84} (2000) 5255}
  [\href{https://arxiv.org/abs/gr-qc/0002040}{{\ttfamily gr-qc/0002040}}].

\bibitem{lqgentropy2}
E.R.~Livine and D.R.~Terno, \emph{{Quantum black holes: Entropy and
  entanglement on the horizon}},
  \href{https://doi.org/10.1016/j.nuclphysb.2006.02.012}{\emph{Nucl. Phys. B}
  {\bfseries 741} (2006) 131}
  [\href{https://arxiv.org/abs/gr-qc/0508085}{{\ttfamily gr-qc/0508085}}].

\bibitem{lqgentropy3}
J.~Engle, K.~Noui, A.~Perez and D.~Pranzetti, \emph{{The SU(2) Black Hole
  entropy revisited}},
  \href{https://doi.org/10.1007/JHEP05(2011)016}{\emph{JHEP} {\bfseries 05}
  (2011) 016} [\href{https://arxiv.org/abs/1103.2723}{{\ttfamily 1103.2723}}].

\bibitem{steffen}
S.~Gielen, \emph{{Group field theory and its cosmology in a matter reference
  frame}}, \href{https://doi.org/10.3390/universe4100103}{\emph{Universe}
  {\bfseries 4} (2018) 103} [\href{https://arxiv.org/abs/1808.10469}{{\ttfamily
  1808.10469}}].

\end{thebibliography}\endgroup
\end{document}